\documentclass[12pt,draftcls,peerreviewca,onecolumn,a4paper,dvips]{IEEEtran}
\usepackage{amsfonts}
\usepackage{mathrsfs}
\usepackage{amssymb}
\usepackage{graphicx}
\usepackage{psfrag}
\usepackage{amsmath}
\usepackage{array}
\usepackage{cases}
\usepackage{color}
\usepackage{url}
\begin{document}
\author{Liangzhong~Ruan,~\IEEEmembership{Student~Member,~IEEE,
}  \\Vincent~K.N.~Lau, \IEEEmembership{Fellow,~IEEE,
}  Xiongbin Rao
\thanks{The authors are with ECE Department, the Hong Kong University of Science and Technology, Hong Kong (e-mails: \{stevenr,eeknlau,xrao\}@ust.hk). This work is funded by RGC 614910.}}

\title{Interference Alignment for Partially Connected MIMO Cellular Networks}

\newtheorem{Thm}{Theorem}[section]
\newtheorem{Lem}{Lemma}[section]
\newtheorem{Lem2}{Lemma}[subsection]
\newtheorem{Asm}{Assumption}[section]
\newtheorem{Def}{Definition}[section]
\newtheorem{Remark}{Remark}[section]
\newtheorem{Prob}{Problem}[section]
\newtheorem{Prob2}{Problem}[subsection]
\newtheorem{Alg}{Algorithm}
\newtheorem{Example}{Example}
\newtheorem{Cor}{Corollary}[section]
%\definecolor{mblue}{rgb}{0.05,0.05,0.6}
\definecolor{mblue}{rgb}{0,0,0}
\maketitle

\begin{abstract}
In this paper, we propose an iterative interference alignment (IA) algorithm for MIMO cellular networks with {\em partial connectivity}, which is induced by heterogeneous path losses and spatial correlation. Such systems impose several key technical challenges in the IA algorithm design, namely the {\em overlapping between the direct and interfering links} due to the MIMO cellular topology as well as how to exploit the {\em partial connectivity}. We shall address these challenges and propose a three stage IA algorithm. As illustration, we analyze the achievable degree of freedom (DoF)
 of the proposed algorithm for a symmetric partially connected MIMO cellular network. We show that there is significant DoF gain compared with conventional IA algorithms due to partial connectivity. The derived DoF bound is also backward compatible with that achieved on fully connected K-pair MIMO interference channels.
\end{abstract}

\section{Introduction}\label{sec:intro}
\begin{figure}[h] \centering
\includegraphics[scale=0.7]{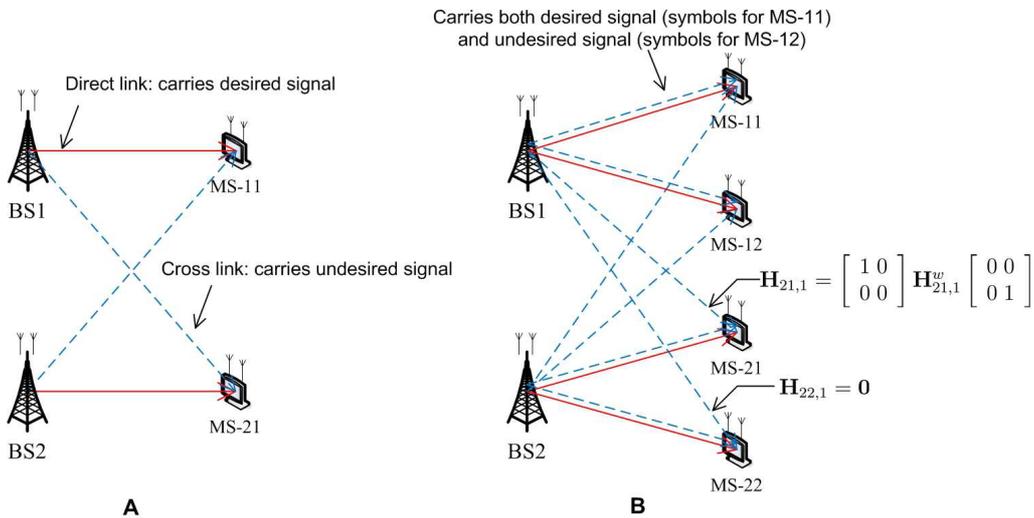}
\caption {Illustration of the cross link - direct link overlapping issue and the partial connectivity in MIMO cellular networks.} \label{fig_KVC}
\end{figure}

Recently, there are intense research interests in the area of interference channels and the associated interference mitigation techniques.  In particular, IA approach can achieve the optimal degree of freedom (DoF) in K-pair interference channels \cite{JafarK1} as well as 2-pair MIMO-X channels \cite{JafarX2}. In \cite{Tse}, the IA approach is extended to cellular OFDMA systems by exploiting some problem-specific structure such as the channel states being full-rank diagonal matrices. In \cite{Tse2, MIMOCell1, MIMOCell2, MIMOCell3}, the authors extend the IA approach to MIMO cellular networks. However, these works have focused on two-cell configuration with one data stream for each mobile (MS) \cite{Tse2,MIMOCell1} or with no more than two MSs in each cell \cite{MIMOCell2,MIMOCell3}, and their extension to general MIMO cellular networks (with arbitrarily number of cells, MSs and data streams) is highly non-trivial. Furthermore, in all these works, a fully connected interference topology is assumed. In practice, we might have heterogeneous path losses between base stations (BSs) and MSs as well as spatial correlation in the MIMO channels. These physical effects induce a {\em partially connected interference topology}.  Intuitively, partial connectivity in interference topology may contribute to limiting the aggregate interference and this may translate into throughput gains in interference-limited systems. Yet, in order to exploit this potential advantage, it is very important to incorporate the partial connectivity topology in the IA algorithm design. In this paper, we are interested to study the potential benefit of partially connectivity in MIMO cellular networks with general configurations and quasi-static fading. There are some key technical challenges that have to be addressed.
\begin{itemize}
\item {\bf Challenges inherent to MIMO cellular networks:}
The existing iterative IA algorithm designed for interference channels \cite{JafarD1} exploit the statistical independency of the direct links  and the cross links.  However, for MIMO cellular networks, there is overlapping between the direct links and the cross links as illustrated in Fig.~\ref{fig_KVC}. As a result, brute force application of the conventional IA schemes in MIMO cellular systems may not have desirable performance.
\item {\bf Challenges to exploit Partial Connectivity:}
In practice, MIMO cellular systems are usually partially connected due to
path losses and spatial correlation, as illustrated in Figure~\ref{fig_KVC}B. Designing an IA algorithm which can exploit the benefit of partial connectivity in the general case is highly non-trivial. While part of this issue has been addressed in our prior work \cite{J_Ruan}, the algorithm proposed in \cite{J_Ruan} cannot be directly extended to the cellular case due to the specific challenges induced by the cellular typology.
\item {\bf Challenges due to Quasi-Static Fading:} For quasi-static interference networks, the IA design may be infeasible \cite{JafarDf}. However, the
IA feasibility checking algorithm proposed in \cite{JafarDf} involves huge complexity of
$\mathcal{O}(2^{N^2})$, where $N$ is the total number of nodes in the network.
Such a complexity is intolerable in practice. Hence, a low
complexity algorithm for checking the IA feasibility conditions on a
real-time basis is needed.
\end{itemize}

In this paper, we will tackle the above challenges by proposing a novel IA algorithm that exploits the {\em partial connectivity topology} in MIMO cellular networks. We adopt an optimization-based approach and decompose the problem into three sub-problems which allows us to tackle the challenges due to MIMO cellular topology and the partial connectivity separately. Moreover, we propose a low complexity IA feasibility checking algorithm that has
worst case complexity of $\mathcal{O}(N^3)$ only. Based on the proposed scheme, we derive an achievable bound of the DoF in a symmetric partially connected MIMO cellular network. We show that using the proposed algorithm, the partial connectivity can be exploited to increase the total DoF in the MIMO cellular networks. Finally, the proposed scheme is compared with various conventional baseline algorithms via simulations and is shown to achieve significant throughput gain.

 The following notations are used in this paper: $a$, $\mathbf{a}$, $\mathbf{A}$, and $\mathbb{A}$ represent scaler, vector, matrix, set/space, respectively, in particular, $\mathbb{R}$, $\mathbb{C}$ represent the set of real number and complex number, respectively. The operators $(\cdot)^T$, $(\cdot)^H$, $\mbox{rank}(\cdot)$, $\mbox{trace}(\cdot)$, {\color{mblue} $|\cdot|$, and $\dim(\cdot)$} denote transpose, hermitian, rank, trace, {\color{mblue} cardinality (of a set) and dimension (of a space),} respectively. $\mbox{span}(\{\mathbf{a}\})$ denotes the linear space spanned by the vectors in $\{\mathbf{a}\}$. $\mbox{span}(\{\mathbf{A}\})$ represents the space spanned by the column vectors of $\mathbf{A}$. $\mathcal{G}(S,N)$ denotes the Grassmannian \cite{J_Grass}, which represents the space of all the $S$ dimensional subspaces of the $N$ dimensional space.

\section{System Model and Problem Formulation}
\subsection{MIMO Cellular Networks}
\label{sec:model} We consider a MIMO cellular system with $G$ BSs, each of which serves $K$ MSs,  as illustrated in Fig.~\ref{fig_KVC}B. Denote $N^t_{g}$, $N^r_{gk}$ as the
number of antennas at BS-$g$ and the $k$-th MS of BS-$g$, respectively. Denote
 $d_{gk}$ as the number of data streams transmitted to the $k$-th MS from BS-$g$. The received signal at the $k$-th MS of BS-$g$ is given by:
\begin{eqnarray}
\mathbf{y}_{gk}=\mathbf{U}_{gk}^{\color{mblue}H}\left(\sum_{n=1}^G
\mathbf{H}_{gk,n}\sum_{i=1}^{K} \mathbf{V}_{ni}\mathbf{x}_{ni}
\;+\mathbf{z}\right),\; \forall k\in\{1,...,K\}
\label{eqn:signal_1}
\end{eqnarray}
where $\mathbf{H}_{gk,n}\in \mathbb{C}^{N^r_{gk}\times N^t_n}$ is the channel state information (CSI) from BS-$n$ to the $k$-th MS of BS-$g$, $\mathbf{x}_{ni}\in\mathbb{C}^{d_{ni}\times 1}$, $\mathbf{V}_{ni}\in\mathbb{C}^{N^t_{n}\times
d_{ni}}$ and $\mathbf{U}_{ni}\in\mathbb{C}^{\color{mblue} N^r_{ni}\times d_{ni}}$ are the information symbols, the precoding matrix and the decorrelator matrix, respectively, for the $i$-th MS of the BS-$g$.
$\mathbf{z}\in\mathbb{C}^{N^r_{gk}\times 1}$ is the white
Gaussian noise with unity variance. The CSI matrices $\{\mathbf{H}_{gk,n}\}$ are assume to be quasi-static and mutually independent random matrices. Furthermore, we normalize the precoding matrix and the transmit symbols as $\mbox{trace}\left({\mathbf{V}^H_{ni}\mathbf{V}_{ni}}\right)=d_{ni}$ and $\mathbb{E}\left[\sum_{i=1}^{K}
\mbox{trace}(\mathbf{x}^H_{ni}\mathbf{x}_{ni})\right] = P_n$ so that the total transmit power from BS-$n$ is $P_n$.

\subsection{Partial Connectivity in MIMO Cellular Networks}
{\color{mblue}We first describe the statistical model of the CSI matrices $\{\mathbf{H}_{gk,n}\}$. This model was first proposed in \cite{J_Channel} and widely adopted in literature.
\begin{Asm}[Channel Fading with Dual-sides Correlation]\label{asm:channel} We consider a channel model that incorporates  both transmit and receive spatial correlation and channel gain, thus:
\begin{eqnarray}
\mathbf{H}_{gk,n}=G_{gk,n}\mathbf{A}^H_{gk,n}\mathbf{H}^w_{gk,n}\mathbf{B}_{gk,n}\label{eqn:channel}
\end{eqnarray}
where $\mathbf{H}^w_{gk,n}\in\mathbb{C}^{N^r_{gk}\times N^t_n}$ contains i.i.d. $\mathcal{CN}(0,1)$  entries, $G_{gk,n}\in \mathbb{R}^+\cup\{0\}$ is the square root of channel gain, $\mathbf{A}_{gk,n}\in\mathbb{C}^{N^r_{gk}\times N^r_{gk}}$, $\mathbf{B}_{gk,n}\in\mathbb{C}^{N^r_{n}\times N^r_{n}}$ represent the receive and transmit spatial correlation, respectively. Here  $\mathbf{A}^H_{gk,n}\mathbf{A}_{gk,n}$, $\mathbf{B}^H_{gk,n}\mathbf{B}_{gk,n}$ are positive semi-definite matrices, $||\mathbf{A}_{gk,n}||_F=||\mathbf{B}_{gk,n}||_F=1$.\hfill~\IEEEQED
\end{Asm}

Based on the statistical model of the CSI matrices, we formally define the notion of {\em partial connectivity} below.
\begin{Def}[Partial Connectivity] We define the partial connectivity between BS-$n$ and the $k$-th MS of BS-$g$ to be the null space of the spatial correlation matrices  $\mathbf{A}_{gk,n}$, $\mathbf{B}_{gk,n}$ time channel gain factor $G_{gk,n}$:
\begin{itemize}
\item{\em Transmit partial connectivity:} $\mathcal{N}^t_{gk,n}\triangleq\mathcal{N}(G_{gk,n}\mathbf{B}_{gk,n})$.
\item{\em Receive partial connectivity:} $\mathcal{N}^r_{gk,n}\triangleq\mathcal{N}(G_{gk,n}\mathbf{A}_{gk,n})$.~\hfill \IEEEQED
\end{itemize}\label{def:model}
\end{Def}
}
\begin{Remark}[Physical Meaning of Partial Connectivity] The partial connectivity actually describes the effective subspaces of the channel matrices between BSs and MSs in the network. For instance, $\{\mathcal{N}^t_{gk,n}, \mathcal{N}^r_{gk,n}\}$ represent the subspaces that cannot be  perceived by the BSs and the MSs, respectively. Hence, the partial connectivity topology of the MIMO cellular network is parameterized by the null spaces  $\{\mathcal{N}^t_{gk,n}, \mathcal{N}^r_{gk,n}\}$. Also note that both the inter-cell links (i.e. $g\neq n$) and the intra-cell links (i.e. $g=n$) may be partially connected.~\hfill~\IEEEQED
\end{Remark}

 We consider a few examples below (as shown in Fig.~\ref{fig_KVC}B) to illustrate how the partial connectivity model in Definition~\ref{def:model} corresponds to various physical situations. Note that CSI matrices $\mathbf{H}_{gk,n}\in\mathbb{C}^{2\times 2}$ are modeled by \eqref{eqn:channel}.
\begin{itemize}
\item{\bf Fully connected MIMO cellular network:} If $G_{gk,n}\neq 0$ and $\mathbf{A}_{gk,n}$, $\mathbf{B}_{gk,n}$ are full rank, we have $\mathcal{N}^t_{gk,n}=\mathcal{N}^r_{gk,n}=\{\mathbf{0}\}$, $\forall g,k$ and this corresponds to a fully connected network.
\item{\bf MIMO cellular network with spatial correlation:} As an illustration, $\mathbf{H}_{21,1}$ has spatial correlation such that $\mathbf{A}_{21,1}=\left[\begin{array}{c@{\quad}c}1&0
    \\0&0\end{array}\right]$, $\mathbf{B}_{21,1}=\left[\begin{array}{c@{\quad}c}0&0
    \\0&1\end{array}\right]$, we have $\mathcal{N}^r_{21,1}=\mbox{span}([0,1]^{\color{mblue}T})$, $\mathcal{N}^t_{21,1}=\mbox{span}([1,0]^{\color{mblue}T})$.
\item{\bf MIMO cellular network with heterogeneous path losses:} Suppose the path loss from BS-$1$ to the second MS of BS-$2$ is 60 dB and the transmit SNR is 40 dB. Since the interference power from BS-$1$ is negligible compared with the gaussian noise, we can effectively assume
$G_{22,1} = {0}$, which gives $\mathcal{N}^t_{22,1}=\mathcal{N}^r_{22,1}=\mathbb{C}^2$, as illustrated in Fig.~\ref{fig_KVC}B.
\end{itemize}

\subsection{Stream Assignment and Transceiver Design under Interference Alignment Constraints}
We assume all the BSs in the MIMO cellular network share global CSI knowledge\footnote{Global CSI is easy to obtain when the network size is small. When the networks size is large, the partial connectivity can be exploited to achieve scalable CSI feedback schemes. For instance, by utilizing heterogeneous path loss, in \cite{J_Feedback}, the authors propose a scalable CSI feedback scheme for MIMO cellular networks.} $\{\mathbf{H}_{gk,n}\}$. We adopt the IA approach to maximize the network total DoF, which is defined by {\color{mblue}$D=\lim_{\mbox{\small SNR}\rightarrow\infty}\frac{C}{\log(\mbox{\small SNR})}$, where $C$ is the network sum throughput and SNR is the signal to noise ratio. Note that  $C=D\log(\mbox{SNR})+${\scriptsize $\mathcal{O}$}$(\log(\mbox{SNR}))$,} DoF gives a first order estimation on network throughput. Moreover, it offers some first order simplification to the complex throughput optimization on MIMO interference network. Specifically, we would like to jointly optimize the data stream
assignment $\{d_{nj}\}$, precoders $\{\mathbf{V}_{nj}\}$
and decorrelators $\{\mathbf{U}_{nj}\}$, $n\in\{1,...,G\},
j\in\{1,...,K\}$ policies to maximize the total number of data streams $\sum_{n=1}^{G}\sum_{k=1}^{K}d_{nj}$ under the IA constraints\footnote{Under the IA constraints \eqref{eqn:rank_1}, \eqref{eqn:cs_cross_1}, the number of data streams equals to the DoF of the network.}, i.e.:
\begin{Prob}[IA for MIMO Cellular Networks]\label{pro:original}
\begin{eqnarray}
&&\max_{\{d_{nj}\},\{\mathbf{V}_{nj}\},\{\mathbf{U}_{gk}\}}\sum_{n=1}^{G}\sum_{j=1}^{K}d_{nj}
\\&\mbox{s.t.:}&\mbox{rank}(\mathbf{U}^H_{gk}\mathbf{H}_{gk,g}
\mathbf{V}_{gk})=d_{gk},
\label{eqn:rank_1}
\\&& \mathbf{U}^H_{gk}\mathbf{H}_{gk,n} \mathbf{V}_{nj}=\mathbf{0},\label{eqn:cs_cross_1}
\\&&\mbox{trace}\left(\mathbf{V}^H_{nj}\mathbf{V}_{nj}\right)=d_{nj},\label{eqn:normv}
\\\nonumber && d_{nj}\in\{0,1,...,d^{\max}_{nj}\},\;\forall g,n\in\{1,...,G\},\;
k,j\in\{1,...,K\},\;(n,j)\neq(g,k)
\end{eqnarray}
where $d^{\max}_{nj}$ is the maximum number of data streams for the concerned MS. Constraint \eqref{eqn:rank_1} ensures that all the direct links have sufficient rank to receive the desired signals while constraint \eqref{eqn:cs_cross_1} ensures that all the undesired signals are aligned.
\end{Prob}
\section{IA for Fully Connected MIMO Cellular Networks}
\label{sec:full}
{\color{mblue}In this section, we shall first solve Problem~\ref{pro:original} for fully connected MIMO cellular networks, i.e. $\dim(\mathcal{N}^t_{gk,n})=\dim(\mathcal{N}^r_{gk,n})=0$.}
\subsection{The Unique Challenge for MIMO Cellular Networks}
In the literature, a common approach towards IA for interference channel is based on the interference leakage minimization iteration \cite{JafarD1}. While this approach is designed for standard interference channels, one can extend the framework to MIMO cellular network as below:
\begin{Alg}[Extension of Existing Iterative IA Algorithm \cite{JafarD1}]\footnote{\color{mblue} The algorithm proposed in \cite{JafarD1} is important as it offers a systematic way to obtain IA transceiver design for MIMO interference networks with {\em general} configuration. Most other existing IA algorithms applies to MIMO interference networks with simple specific configuration only. Please refer to \cite[Sec. I]{J_Ruan} for details.}\label{pro:naive}
Alternatively update precoders $\mathbf{V}_{nj}$ and decorrelators
$\mathbf{U}_{gk}$ by minimizing the total interference leakage expressions in \eqref{eqn:optvc} and \eqref{eqn:optuc}  until the algorithm converges.
\begin{eqnarray}&&\min_{\mathbf{V}_{nj}\in\mathbb{C}^{N^t_n\times d_{nj}}\atop
\mathbf{V}^H_{nj}\mathbf{V}_{nj}=\mathbf{I}}\sum^G_{g=1}\sum_{k=1\atop{(g,k)\neq (n,j)}}^{K}\mbox{trace}\left((\mathbf{U}^H_{gk}\mathbf{H}_{gk,n}
\mathbf{V}_{nj})^H(\mathbf{U}^H_{gk}\mathbf{H}_{gk,n}
\mathbf{V}_{nj})\right)\label{eqn:optvc}
\\&&\min_{\mathbf{U}_{gk}\in\mathbb{C}^{N^r_{gk} \times d_{gk}}\atop
\mathbf{U}^H_{gk}\mathbf{U}_{gk}=\mathbf{I}}\sum_{n=1}^G\sum_{j=1\atop (n,j)\neq (g,k)}^{K}\mbox{trace}\left((\mathbf{U}^H_{gk}\mathbf{H}_{gk,n}
\mathbf{V}_{nj})^H(\mathbf{U}^H_{gk}\mathbf{H}_{gk,n}
\mathbf{V}_{nj})\right)\label{eqn:optuc}
\end{eqnarray}
$\forall n,g\in\{1,...,G\}$, $j,k\in\{1,...,K\}$.\end{Alg}

\begin{figure} \centering
\includegraphics[scale=0.63]{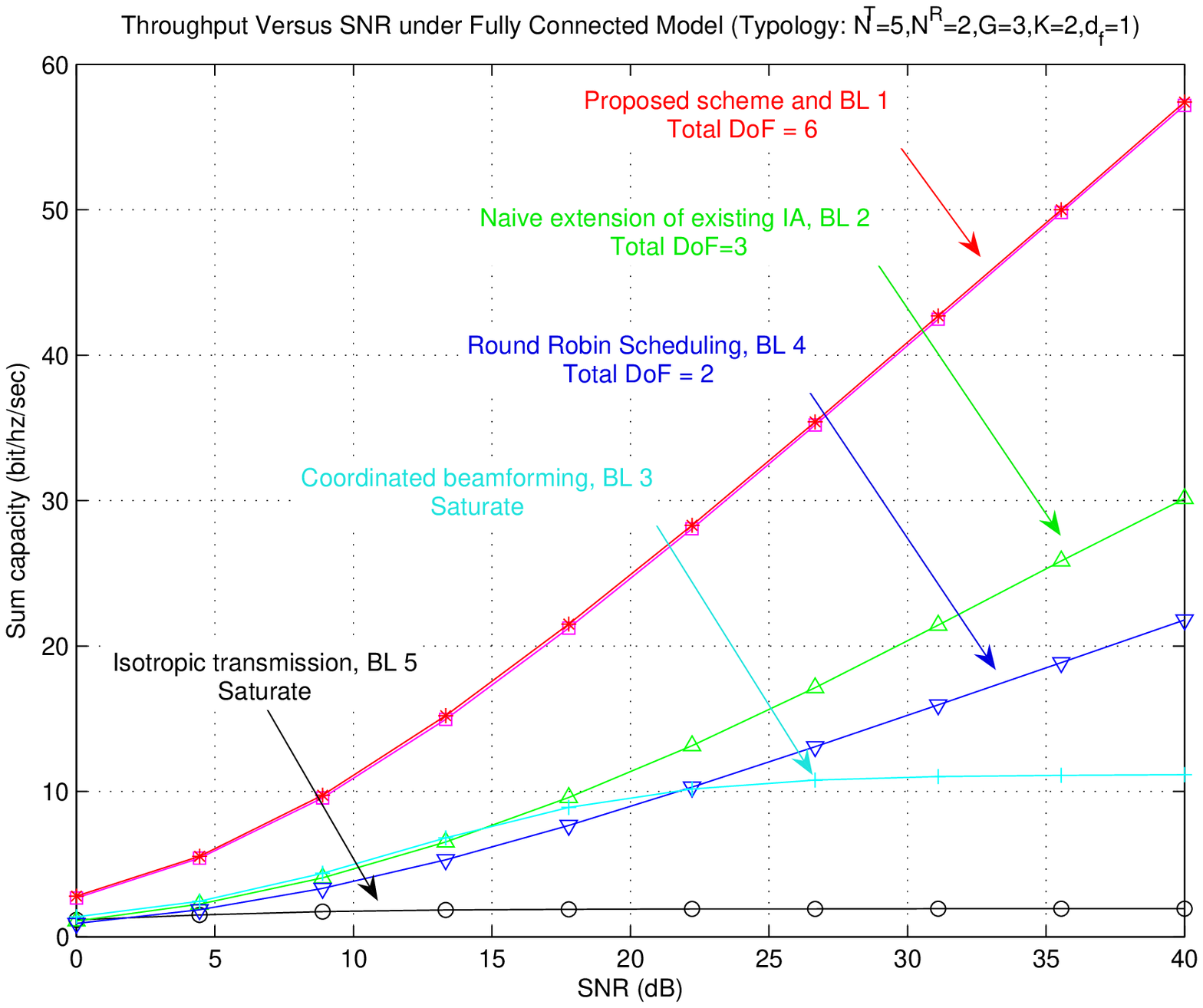}
\caption {Sum throughput versus SNR for the proposed algorithm (and
5 baselines) in a fully connected MIMO cellular network.} \label{fig_fullyconnected}
\end{figure}

Fig.~\ref{fig_fullyconnected} illustrates the performance of the naive algorithm in a 3-BS fully connected MIMO cellular network with $K=2$, $N^t_g=5$, $N^r_{gk}=2$, $d_{gk}=1$, $\forall g\in\{1,3\}, k\in\{1,2\}$. It is shown that the naive algorithm could achieve a total DoF of 3, which is only half of the achievable DoF lower bound given in \cite{JafarDf}, which demonstrates that naive extension of standard iterative IA algorithm can perform poorly in MIMO cellular networks. This problem is due to the direct link - cross link overlapping issue defined below:
\begin{Def}[Direct Link - Cross Link Overlapping] In an interference network, denote the set of the channels that carry the desired signals and undesired signals as $\mathbb{H}^{D}$ and $\mathbb{H}^{C}$, respectively. If $\mathbb{H}^{D}\cap\mathbb{H}^{C}
\neq \emptyset$, then the network has direct link - cross link overlapping.~\hfill~\IEEEQED
\end{Def}

As illustrated in Fig.~\ref{fig_KVC}A, in conventional MIMO interference network, $\mathbb{H}^D=\{\mathbf{H}_{mm}\}$ and $\mathbb{H}^C=\{\mathbf{H}_{mn}:m\neq n\}$, where $m$, $n$ are the indexes for transmitters and receivers, respectively. Obviously, in this case, $\mathbb{H}^{D}\cap\mathbb{H}^{C}
= \emptyset$ and there is no overlapping issue. However, as illustrated in Fig.~\ref{fig_KVC}B, in MIMO cellular network, when the number of MSs per cell $K>1$, the intra-cell links also carries over undesired signals and hence $\mathbb{H}^D=\{\mathbf{H}_{gk,g}\}$ and $\mathbb{H}^C=\{\mathbf{H}_{gk,n}\}$. In this scenario, we have that $\mathbb{H}^{D}\cap\mathbb{H}^{C}=\mathbb{H}^{D}
\neq \emptyset$.  As the channel states in $\mathbb{H}^D$ appear in \eqref{eqn:optvc} and \eqref{eqn:optuc}, when we update the precoders and decorrelators via \eqref{eqn:optvc} and \eqref{eqn:optuc}, we may also reduce the dimension of the signal space for the desired signals as well.
\subsection{Problem Decomposition}
We
decompose the original problem, i.e. Problem~\ref{pro:original} into the following three subproblems:
\begin{Prob}[Stream Assignment]\label{pro:user}
\begin{eqnarray}&&\max_{\{d_{nj}\}}\sum_{n=1}^{G}\sum_{j=1}^Kd_{nj}
\\&\mbox{S.t.}&\sum_{(g,k)\in\mathbb{S}_U, (n,j)\in\mathbb{S}_V\atop {g\neq n} }d_{gk}d_{nj} \le
\sum_{(n,j)\in\mathbb{S}_V}d_{nj}(N^t_n-{\sum_{k=1}^K
d_{nk}})+\sum_{(g,k)\in\mathbb{S}_U}d_{gk}(N^r_{gk}-d_{gk})\label{eqn:feasible1}
\\&&\nonumber \forall \mathbb{S}_V, \mathbb{S}_U \subseteq
\mathbb{S}, \mbox{ where } \mathbb{S}=\{(g,k):g\in\{1,...,G\},
k\in\{1,...,K\}\}
\end{eqnarray}
\end{Prob}
\begin{Prob}[Inter-cell Interference Suppression]\label{pro:inter}
\begin{eqnarray}
&&\min_{\mathbf{V}^F_{nj},\mathbf{U}_{gk}}\sum_{g=1\atop \neq
n}^{G}\sum_{k=1}^{K}\mbox{trace}\left((\mathbf{U}^H_{gk}\mathbf{H}_{gk,n}
\mathbf{V}^I_{nj})^H(\mathbf{U}^H_{gk}\mathbf{H}_{gk,n}
\mathbf{V}^I_{nj})\right) \label{eqn:inter}
\\ &\mbox{S.t. }&
\mathbf{V}^I_{nj}=\mathbf{V}^C_{nj}+\mathbf{S}_n\mathbf{V}^F_{nj},
\;\mathbf{V}^F_{nj}\in\mathbb{C}^{(N^t_n-\sum_{k=1}^K
d^*_{nk})\times d^*_{nj}} \label{eqn:vform_1}
\\&&  \mathbf{U}^H_{gk}\mathbf{U}_{gk}=\mathbf{I},\mathbf{U}_{gk}\in\mathbb{C}^{N^r_{gk}\times
d_k}\label{eqn:uform_1}
\end{eqnarray}
$\forall g\in\{1,...,G\},k\in\{1,...,K\}$, where:
$\{{d}^*_{nj}\}$ are the solutions of Problem~\ref{pro:user}, matrices
$\mathbf{V}^C_{nj}\in\mathbb{C}^{N^t_n\times d^*_{nj}}$,
$\mathbf{S}_n\in\mathbb{C}^{N^t_n \times
(N^t_n-\sum_{k=1}^K d^*_{nk})}$, are isometry matrices
whose row vectors combined together form a basis for
$\mathbb{C}^{N^t_n\times 1}$, i.e.
\begin{eqnarray}\left[\mathbf{V}^C_{n1},
\mathbf{V}^C_{n2},...\mathbf{V}^C_{nK},\mathbf{S}_n\right]^H\left[\mathbf{V}^C_{n1},
\mathbf{V}^C_{n2},...\mathbf{V}^C_{nK},\mathbf{S}_n\right]=\mathbf{I}.\label{eqn:orth}\end{eqnarray}
\end{Prob}
\begin{Prob}[Intra-cell Interference Suppression]\label{pro:intra}
\begin{eqnarray}
&&\min_{\mathbf{V}_{nj}}\sum_{k=1,\neq
j}^K\mbox{trace}\left(((\mathbf{U}^{*}_{nk})^H\mathbf{H}_{nk,n}\mathbf{V}_{nj})^H
((\mathbf{U}^{*}_{nk})^H\mathbf{H}_{nk,n}\mathbf{V}_{nj})\right)\label{eqn:intra}
\\&\mbox{S.t.}&
(\mathbf{V}_{nj})^H\mathbf{V}_{nj}=\mathbf{I}\label{eqn:viso}
\\&&\mbox{rank}\left((\mathbf{U}^*_{nj})^H\mathbf{H}_{nj,j}\mathbf{V}_{nj}\right)=d^*_{nj}\label{eqn:uvrank2}
\\&&\mbox{span}\left(\left[\mathbf{V}_{n1},\mathbf{V}_{n2},...,\mathbf{V}_{nK}\right]\right)
\subseteq\mbox{span}\left(\left[\mathbf{V}^{I*}_{n1},\mathbf{V}^{I*}_{n2},...,\mathbf{V}^{I*}_{nK}\right]\right)
\label{eqn:spaceeq}
\end{eqnarray}
where $\{\mathbf{V}^{I*}_{nj}\}$ and $\{\mathbf{U}^*_{nj}\}$,
$n\in\{1,...,G\},j\in\{1,...,K\}$, are the solutions of
Problem~\ref{pro:inter}, $\mbox{span}(\mathbf{X})$ denotes the
linear space spanned by the row vectors of $\mathbf{X}$.~\hfill
\IEEEQED
\end{Prob}

\begin{Thm}[Equivalence between the Original Problem and the Subproblems]
For fully connected MIMO cellular networks with i.i.d. channel matrices $\{\mathbf{H}_{gk,n}\}$, the optimizing variables of Problem~\ref{pro:original} is given by ($\{d^*_{gk}\}$, $\{\mathbf{U}^*_{gk}\}$, $\{\mathbf{V}^*_{gk}\}$) with probability 1, where $\{d^*_{gk}\}$, $\{\mathbf{U}^*_{gk}\}$, $\{\mathbf{V}^*_{gk}\}$ are the solutions of Subproblem~\ref{pro:user},~\ref{pro:inter},~\ref{pro:intra} respectively. Furthermore, the optimal value of Problem 2.1 is $D^*=\sum_{n=1}^{G}\sum_{j=1}^{K}d^*_{nj}$.\label{thm:equal}
\end{Thm}
\proof Please refer to Appendix\ref{pf_thm:equal} for the proof.
\endproof

\begin{Remark}[Roles of the Three Subproblems]
\begin{itemize}
\item[]
\item\emph{Problem~\ref{pro:user}} determines the stream assignment
$\{d_{nj}\}$ to maximize the sum of the data stream
numbers (i.e. DoF) of the network, conditioned on the network being IA
feasible.
\item \emph{Problem~\ref{pro:inter}} updates the intermediate precoders
$\{\mathbf{V}^I_{nj}\}$ and decorrelators $\{\mathbf{U}_{gk}\}$
to suppress the inter-cell interferences.
\item \emph{Problem~\ref{pro:intra}} further adjusts the precoders
$\{\mathbf{V}_{nj}\}$ to suppress the intra-cell interferences.
\end{itemize}

Note
that after separating the process of inter-cell and intra-cell
interference mitigation in Problem~\ref{pro:inter} and Problem~\ref{pro:intra}, only inter-cell channel states
$\{\mathbf{H}_{gk,n}\}$, $g\neq n$ are involved in Problem~\ref{pro:inter}. This property is very important to overcome the cross link - direct link
overlapping issue.~\hfill \IEEEQED
\end{Remark}

\begin{figure} \centering
\includegraphics[scale=0.6]{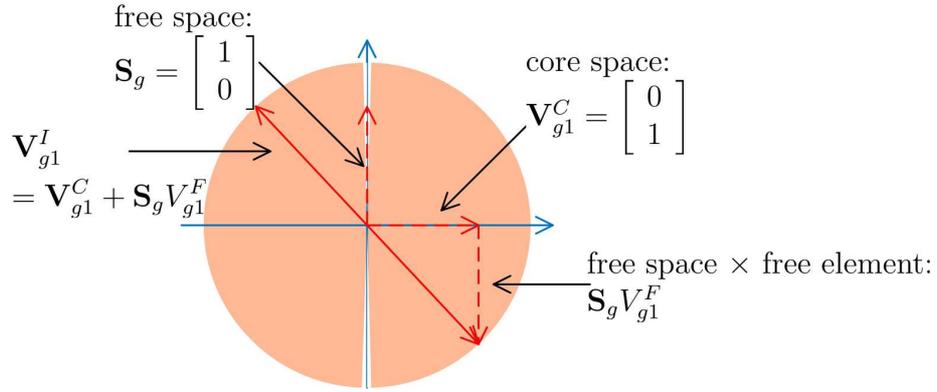}
\caption {A simple example of the Core space, Free space and
Free elements. In this figure, $N^t_g=2$, $d_{gk}=1$.} \label{fig_Vform}
\end{figure}

\begin{Remark}[The Structure of the Intermediate Precoder]\label{remark:three} Unlike the existing iterative IA algorithm, we have introduced an auxiliary variable, namely the intermediate precoder variables $\{\mathbf{V}^I_{nj}\}$. From \eqref{eqn:vform_1}, $\mathbf{V}^I_{nj}$  consists of the {\em core space} $\mathbf{V}^C_{nj}$, the {\em free space} $\mathbf{S}_{n}$ and the free elements $\mathbf{V}^F_{nj}$ as illustrated in Fig.~\ref{fig_Vform}. This precoder structure in the auxiliary variable  enables us to separate inter-cell and
intra-cell interference suppression.
~\hfill \IEEEQED
\end{Remark}

\begin{Remark}[The Physical Meaning of Equation \eqref{eqn:spaceeq}]
Constraint \eqref{eqn:spaceeq} is introduced to make sure that the desirable  \emph{inter-cell} interference alignment property obtained
in Problem~\ref{pro:inter} is still maintained during the precoder updates $\{\mathbf{V}_{nj}\}$ in Problem~\ref{pro:intra}. This is because of the following. Suppose $\{\mathbf{U}^*_{gk}\}$ and $\{\mathbf{V}^{I*}_{nj}\}$ constitute the solution of Problem~\ref{pro:inter}. We have
$(\mathbf{U}^*_{gk})^H\mathbf{H}_{gk,n}\left[\mathbf{V}^{I*}_{n1},
\mathbf{V}^{I*}_{n2},...,\mathbf{V}^{I*}_{nK}\right]=\mathbf{0}$,
$\forall g\neq n\in\{1,...,G\}$. From \eqref{eqn:spaceeq}, there must exist a matrix
$\mathbf{R}_n\in\mathbb{C}^{\sum_{j=1}^Kd_{nj}\times\sum_{j=1}^Kd_{nj}}$
such that
$\left[\mathbf{V}_{n1},\mathbf{V}_{n2},...,\mathbf{V}_{nK}\right]=\left[\mathbf{V}^{I*}_{n1},
\mathbf{V}^{I*}_{n2},...,\mathbf{V}^{I*}_{nK}\right]$
$\mathbf{R}_n$, which leads to the following equation
\begin{eqnarray}
(\mathbf{U}^*_{gk})^H\mathbf{H}_{gk,n}\left[\mathbf{V}_{n1},\mathbf{V}_{n2},...,\mathbf{V}_{nK}\right]
=(\mathbf{U}^*_{gk})^H\mathbf{H}_{gk,n}\left[\mathbf{V}^{I*}_{n1},
\mathbf{V}^{I*}_{n2},...,\mathbf{V}^{I*}_{nK}\right]\mathbf{R}_n=
\mathbf{0}\cdot\mathbf{R}_n=\mathbf{0}.\label{eqn:inter_1}
\end{eqnarray}

Hence, equation \eqref{eqn:inter_1} shows that the inter-cell interference
alignment property is preserved for the updated precoders
$\{\mathbf{V}_{nj}\}$ in Problem~\ref{pro:intra}.~\hfill \IEEEQED
\end{Remark}

\subsection{Solution to Problem~\ref{pro:user} (Stream Assignment Problem)}
\label{sec:user}Problem~\ref{pro:user} is a combinatorial problem
whose optimal solution $\{d^*_{gk}\}$
often involves exhaustive search with exponential complexity w.r.t.
to the total number of MSs $GK$.
 For low complexity consideration, we propose the following greedy-based solution.
\begin{Alg}[Greedy Stream Assignment]
\label{alg:offline_f}
\begin{itemize}
\item[]
\item{\bf Step 1 Initialization:} Initialize the stream assignment policy to be
the number of streams requested by each MSs, i.e. $d_{gk}=d^{\max}_{gk}$, $\forall g\in\{1,...,G\}, k\in\{1,...,K\}$.
\item{\bf Step 2 Low complexity IA feasibility checking:}
\begin{itemize}
\item Denote $v^t_{nj},v^r_{gk}$, $n,g\in\{1,...,G\}$,
$j,k\in\{1,...,K\}$ as the number of the freedoms, i.e. free variables
in precoder $\mathbf{V}^I_{nj}$ and decorrelator $\mathbf{U}_{gk}$,
respectively. Note that the number of freedoms in $\mathbf{V}^I_{nj}$ are given by the number of elements in $\mathbf{V}^F_{nj}$ and that in $\mathbf{U}_{gk}$ are given by the dimension of Grassmannian $\mathcal{G}(d_{gk},N^r_{gk})$, we have \begin{eqnarray}v^t_{nj}=d_{nj}(N^t_g-\sum_{k=1}^{K}d_{nk}),
v^r_{gk}=d_{gk}(N^r_{gk}-d_{gk})\label{eqn:variable}.\end{eqnarray}

\item Denote $c_{gk,nj}$, $n,g\in\{1,...,G\}$, $j\in\{1,...,K\}$,
$k\in\{1,...,K\}$, as the number of constraints
required to eliminate the interference from $\mathbf{V}_{nj}$ to
$\mathbf{U}_{gk}$. Set
\begin{eqnarray}c_{gk,nj}=d_{nj}d_{gk}, \mbox{ if } g\neq n;\;c_{gk,nj}=0, \mbox{ otherwise.}\label{eqn:constraint}\end{eqnarray}

\item Use the low complexity IA feasibility checking algorithm proposed in Appendix\ref{alg_detail} to check if the system is IA feasible.
 If the network is not IA feasible, go to
Step 3. Otherwise, let $d^*_{gk}=d_{gk}$,
$g\in\{1,...,G\}$, $k\in\{1,...,K\}$ and exit the algorithm.
\end{itemize}
\item{\bf Step 3 :}
 Update $d_{g'k'}=d_{g'k'}-1$ and go back to Step
2, where $(g',k')$ is given by
\begin{eqnarray}
\nonumber (g',k')&=&\arg\max_{g,k}\left(\sum_{n=1}^{G}\sum_{j=1
}^K(c_{gk,nj}+c_{nj,gk}-c'_{gk,nj}-c'_{nj,gk})-(v^t_{gk}+v^r_{gk}-{v'}^t_{gk}-{v'}^r_{gk})\right)
\\&=&\arg\max_{g,k}\left(2\sum_{n=1}^{G}\sum_{j=1\atop(n,j)\neq(g,k)
}^Kd_{nj}-(N^t_{g}+N^r_{gk}-4d_{gk}+2) \right)\label{eqn:selectn}\end{eqnarray}
where $\{{v'}^t_{gk},{v'}^r_{gk}\}$, and $\{c'_{gk,nj},c'_{nj,gk}\}$ denote the number of freedoms and constraints given by \eqref{eqn:variable} and \eqref{eqn:constraint}, respectively, with $d'_{gk}=d_{gk}=1$.
\end{itemize}
\end{Alg}

\begin{Thm}[Property of the Low Complexity IA
Feasibility Checking] The IA feasibility constraint in \eqref{eqn:feasible1} is satisfied if and only if it can satisfy the
low complexity IA feasibility checking in Appendix\ref{alg_detail}. Moreover, the worst case complexity
of the proposed checking scheme is
$\mathcal{O}(G^3K^3)$, which is substantially lower compared with the complexity $\mathcal{O}(2^{G^2K^2})$ in conventional IA feasibility checking \cite{JafarDf}.\label{thm:proper}
\end{Thm}
\proof
Please refer to \cite{Ruan} for the proof.
\endproof

\subsection{Solution to Problem~\ref{pro:inter} (Inter-cell Interference Suppression Problem)}
The following algorithm solves Problem~\ref{pro:inter} by alternatively updating the intermediate precoders $\{\mathbf{V}^I_{nj}\}$ and the decorrelators $\{\mathbf{U}_{gk}\}$ to minimize the inter-cell interference, i.e.:
\begin{eqnarray}
\min_{\mathbf{V}^F_{nj}}\sum_{g=1\atop \neq
n}^{G}\sum_{k=1}^{K}\mbox{trace}\left((\mathbf{U}^H_{gk}\mathbf{H}_{gk,n}
\mathbf{V}^I_{nj})^H(\mathbf{U}^H_{gk}\mathbf{H}_{gk,n}
\mathbf{V}^I_{nj})\right),&\mbox{S.t.:}& \mbox{equation } \eqref{eqn:vform_1},\label{eqn:optvc2}
\\\min_{\mathbf{U}_{gk}}\sum_{g=1\atop \neq
n}^{G}\sum_{k=1}^{K}\mbox{trace}\left((\mathbf{U}^H_{gk}\mathbf{H}_{gk,n}
\mathbf{V}^I_{nj})^H(\mathbf{U}^H_{gk}\mathbf{H}_{gk,n}
\mathbf{V}^I_{nj})\right),&\mbox{S.t.:}& \mbox{equation } \eqref{eqn:uform_1}.\label{eqn:optuc2}
\end{eqnarray}
\begin{Alg}[Alternative Inter-cell Interference Suppression]\label{alg:inter_full}
\begin{itemize}
\item[]
\item{\bf Step 1 Initialization :} Randomly generate
$\mathbf{V}^F_{nj}$, $\forall n\in\{1,...,G\}$,
$j\in\{1,...,K\}$.
\item{\bf Step 2 Minimize interference leakage at the receiver
side:} At the $k$-th MS of $\mbox{BS-}g$ , update $\mathbf{U}_{gk}$:
$\mathbf{u}_{gk}(d)=\nu_d\left[\sum_{n=1,\neq
g}^{G}\sum_{j=1}^KP_{nj}(\mathbf{H}_{gk,n}
\mathbf{V}^I_{nj})(\mathbf{H}_{gk,n}
\mathbf{V}^I_{nj})^H\right]$, where $\mathbf{u}_{gk}(d)$ is
the $d$-th column of $\mathbf{U}_{gk}$, $\nu_d[\mathbf{A}]$ is the
eigenvector corresponding to the $d$-th smallest eigenvalue of
$\mathbf{A}$, $d\in\{1,...,d_{gk}\}$.
\item{\bf Step 3 Minimize interference leakage at the transmitter side:}
At BS-$n$, update $\mathbf{V}^F_{nj}$, $j\in\{1,...,K\}$:
$\mathbf{V}^F_{nj}=-(\mathbf{S}^H_{n}\mathbf{Q}_{nj}\mathbf{S}_{n})^{-1}\mathbf{S}^H_{n}
\mathbf{Q}_{nj}\mathbf{V}^C_{nj}$ ,where
$\mathbf{Q}_{nj}=\sum_{g=1,\neq
n}^{G}\sum_{k=1}^K$ $P_{nj}(\mathbf{U}^H_{gk}\mathbf{H}_{gk,n}
)^H(\mathbf{U}^H_{gk}\mathbf{H}_{gk,n} )$.
\item Repeat Step 2 and 3 until $\mathbf{V}^F_{nj}$ and $\mathbf{U}_{gk}$
converges. Set
$\mathbf{V}^{I*}_{nj}=\mathbf{V}^C_{nj}+\mathbf{S}_n\mathbf{V}^{F}_{nj}$
and $\mathbf{U}^*_{gk}=\mathbf{U}_{gk}$.
\end{itemize}
\end{Alg}
\begin{Thm}[Convergence of Algorithm~\ref{alg:inter_full}]
For fully connected MIMO cellular network with i.i.d. channel matrices $\{\mathbf{H}_{gk,n}\}$, Algorithm~\ref{alg:inter_full} converges to a local optimal solution of Problem~\ref{pro:inter}. Note that global optimality is not assured.\label{thm:inter_full}
\end{Thm}
\proof Please refer to Appendix\ref{pf_thm:inter_full} for the
proof.
\endproof
\begin{Thm}[Property of $\{\mathbf{V}^{I*}_{nj}\}$ and  $\{\mathbf{U}^*_{nj}\}$]
For fully connected MIMO cellular network with i.i.d. channel matrices $\{\mathbf{H}_{gk,n}\}$, the converged solution of Algorithm~\ref{alg:inter_full}
$\{\mathbf{V}^{I*}_{nj}\}$, $\mathbf{U}^*_{nj}$,
$n\in\{1,...,G\}$, $j\in\{1,...,K\}$, satisfy
\begin{eqnarray}
\mbox{rank}\left(\left[\begin{array}{c}(\mathbf{U}^*_{n1})^H\mathbf{H}_{n1,n}
\\(\mathbf{U}^*_{n2})^H\mathbf{H}_{n2,n}
\\\dotfill
\\(\mathbf{U}^*_{nK})^H\mathbf{H}_{nK,n}\end{array}\right]
\left[\mathbf{V}^{I*}_{n1},\mathbf{V}^{I*}_{n2},...,\mathbf{V}^{I*}_{nK}\right]\right)
&=&\sum_{j=1}^Kd_{nj}, \;\forall n\in\{1,..,G\}
\label{eqn:sufficient_rank}
\end{eqnarray}
almost surely.\label{thm:drank}
\end{Thm}
\proof Please refer to Appendix\ref{pf_thm:drank} for the proof.
 \endproof

\subsection{Solution to Problem~\ref{pro:intra} (Intra-cell Interference Suppression Problem)}
We solve Problem~\ref{pro:intra} by the following constructive
algorithm.

\begin{Alg} [Intra-cell Zero-Forcing]\label{alg:intra_full}  Denote
$\overline{\mathbf{W}}_q=\left[\begin{array}{c}\mathbf{W}^H_1,..., \mathbf{W}^H_{q-1}, \mathbf{W}^H_{q+1},...,\mathbf{W}^H_K,\mathbf{W}^H_q
\end{array}\right]^H$, where $\mathbf{W}_q=(\mathbf{U}^*_{nq})^H\mathbf{H}_{nq,n}$, $q\in\{1,...,K\}$. Each BS does the following for
every $q\in\{1,...,K\}$ to calculate the precoders:
\begin{itemize}
\item{\bf Step 1: } Perform LQ decomposition for
$\overline{\mathbf{W}}_q
\left[\mathbf{V}^{I*}_{n1},\mathbf{V}^{I*}_{n2},...,
\mathbf{V}^{I*}_{nK}\right]=\mathbf{L}_{n}(q)\mathbf{Q}_{n}(q)$,
where $\mathbf{Q}_{n}(q)$ is an
$\sum_{j=1}^Kd_{nj}\times
\sum_{j=1}^Kd_{nj}$ unitary matrix, and
$\mathbf{L}_{n}(q)$ is a $\sum_{j=1}^Kd_{nj}\times
\sum_{j=1}^Kd_{nj}$ lower triangular matrix.
\item {\bf Step 2: } Set $\mathbf{V}'_{nq}=\left[\mathbf{V}^{I*}_{n1},\mathbf{V}^{I*}_{n2},...,
\mathbf{V}^{I*}_{nK}\right]\mathbf{Q}'_{n}(q)$, where
$\mathbf{Q}'_{n}(q)$ is a matrix aggregated by the last $d_{nq}$
columns of $\mathbf{Q}^H_{n}(q)$.
\item {\bf Step 3: } Perform singular value decomposition for $\mathbf{V}'_{nq}=\mathbf{A}_{nq}\mathbf{S}_{nq}\mathbf{B}^H_{nq}$,
where $\mathbf{S}_{nq}$ is a $N^t_g\times d_{nq}$ matrix,
$\mathbf{A}_{nq}$ and $\mathbf{B}_{nq}$ are $N^t_g\times
N^t_g$ and $d_{nq}\times d_{nq}$ matrices, respectively. Set
$\mathbf{V}^*_{nq}=\mathbf{A}'_{nq}$, where $\mathbf{A}'_{nq}$
is a matrix aggregated by the first $d_{nq}$ columns of
$\mathbf{A}_{nq}$.
\end{itemize}
\end{Alg}
\begin{Thm}[Optimality of $\{\mathbf{V}^*_{nj}\}$]
For fully connected MIMO cellular network with i.i.d. channel matrices $\{\mathbf{H}_{gk,n}\}$, the output of Algorithm~\ref{alg:intra_full}
$\{\mathbf{V}^*_{nj}\}$, $n\in\{1,...,G\}$, $j\in\{1,...,K\}$, is
the optimal solution for Problem~\ref{pro:intra} almost surely (with
optimal value (intra-cell interference power) $=0$).\label{thm:intra_full}
\end{Thm}
\proof Please refer to Appendix\ref{pf_thm:intra_full} for the
proof.
\endproof
\section{IA for MIMO Cellular Networks with Partial Connectivity}\label{sec:partial}
  \subsection{Space Restriction on Transceivers}
  In the prior work \cite{J_Ruan}, we have shown that in contrast to the conventional cases, partial connectivity can be beneficial to system performance in MIMO interference networks as it gives us an extra dimension of freedom, namely the \emph{interference nulling} to eliminate interference\footnote{Please refer to \cite[Section III-A]{J_Ruan} for detailed elaboration on the concept of interference nulling and the difficulties to integrate it into IA processing. We shall omit the details here due to page limitation.}. In particular, we have found by restricting transceivers to lower dimensional subspaces in partially connected MIMO interference network, we can eliminate many IA constraints at a cost of only a few freedoms in transceiver design and hence extend the IA feasibility region. We exploit the idea of subspace constraint to extend the approach in Section~\ref{sec:full} to exploit the partial connectivity in MIMO cellular networks. Specifically, we impose the following structure on transceivers:
  \begin{Def}[Transceiver Structure to Exploit Partial Connectivity]\label{def:trans_partial}
  \begin{itemize}
  \item[]
  \item{\bf Intermediate precoder with dynamic free space:} $\mathbf{V}^I_{nj}=\mathbf{V}^C_{nj}+\mathbf{S}^t_{nj}\mathbf{V}^F_{nj}$,
  \item{\bf Decorrelator with dynamic linear filter:} $\mathbf{U}_{nj}=\mathbf{S}^r_{nj}\mathbf{U}^F_{nj}$,
  \end{itemize}
  where $\mathbf{S}^t_{nj}\in\mathbb{C}^{N^t_n\times S^t_{nj}}$, $\mathbf{V}^F_{nj}\in\mathbb{C}^{S^t_{nj}\times d_{nj}}$, $\mathbf{S}^r_{nj}\in\mathbb{C}^{N^r_{nj}\times(d_{nj}+S^r_{nj})}$,
  $\mathbf{U}^F_{nj}\in\mathbb{C}^{(d_{nj}+S^r_{nj})\times d_{nj}}$, $S^t_{nj}\in\{0,1,...N^t_n-\sum_{k=1}^Kd_{nj}\}$, $S^r_{nj}\in\{0,1,...N^r_{nj}-d_{nj}\}$.~\hfill~\IEEEQED
  \end{Def}
  \begin{Remark}[Space Restriction via New Transceiver Structures]
    Note that $\mbox{span}(\mathbf{V}^I_{nj})\subseteq \mbox{span}(\mathbf{V}^C_{nj})+\mbox{span}(\mathbf{S}^t_{nj})$, and $\mbox{span}(\mathbf{U}_{nj})\subseteq \mbox{span}(\mathbf{S}^r_{nj})$, space restriction is imposed on $\mathbf{V}^I_{nj}$ and $\mathbf{U}_{nj}$ by the new transceiver structure. As a special case, when $S^t_{nj}=N^t_n-\sum_{k=1}^Kd_{nj}$, $S^r_{nj}=N^r_{nj}-d_{nj}$, the transceiver structure is reduced to that proposed in Section~\ref{sec:full}.~\hfill \IEEEQED
  \end{Remark}

\subsection{Problem Decomposition}
Similar to Section~\ref{sec:full}, the original Problem~\ref{pro:original} is decomposed into three subproblems. The data stream assignment subproblem is modified as below.
\begin{Prob}[Stream Assignment and Subspaces Design]\label{pro:userspace}
\begin{eqnarray}&&\max_{\{d_{nj}\},\{\mathbf{V}^C_{nj}\},\{\mathbf{S}^t_{nj}\},\{\mathbf{S}^r_{nj}\}  }
\sum_{n=1}^{G}\sum_{j=1}^Kd_{nj}
\\&\mbox{S.t.}&\nonumber\sum_{(g,k)\in\mathbb{S}_U,\atop (n,j)
\in\mathbb{S}_V, g\neq n
}\min\left(d_{gk},\dim\left((\mbox{span}(\mathbf{S}^r_{gk})\cap(\mathcal{N}^r_{gk,n})^\bot\right)
\right)\min\left(d_{nj},
\right.
\\&&\left.\dim\left((\mbox{span}(\mathbf{V}^C_{nj})+
\mbox{span}(\mathbf{S}^t_{nj}))\cap(\mathcal{N}^t_{gk,n})^\bot\right)
\right)\le
\sum_{(n,j)\in\mathbb{S}_V}d_{nj}S^t_{nj}+\sum_{(g,k)\in\mathbb{S}_U}d_{gk}S^r_{nj},
\label{eqn:feasible_p}
\\&&\nonumber \forall \mathbb{S}_V, \mathbb{S}_U \subseteq
\mathbb{S}, \mbox{ where } \mathbb{S}=\{(g,k):g\in\{1,...,G\},
k\in\{1,...,K\}\}
\\&&\left[\mathbf{V}^C_{n1},\mathbf{V}^C_{n2},...,\mathbf{V}^C_{nK},\mathbf{S}^t_{nj}\right]^H
\left[\mathbf{V}^C_{n1},\mathbf{V}^C_{n2},...,\mathbf{V}^C_{nK},\mathbf{S}^t_{nj}\right]=\mathbf{I},\; \mathbf{V}^C_{nj}\in(\mathcal{N}^t_{nj,n})^\perp,\label{eqn:orth_p}
\\&&(\mathbf{S}^r_{nj})^H\mathbf{S}^r_{nj}=\mathbf{I}  ,\; \mathbf{S}^r_{nj}\in(\mathcal{N}^r_{nj,n})^\perp, \label{eqn:SR}\; \forall
n\in\{1,...,G\}, j\in\{1,...,K\}.
\end{eqnarray}
\end{Prob}

The second and third subproblems are similar to Problem~\ref{pro:inter} and~\ref{pro:intra} except  replacing
$\mathbf{V}^C_{nj}$, $\mathbf{S}_n$, and $\mathbf{U}_{nj}$ with $\mathbf{V}^{C*}_{nj}$,
$\mathbf{S}^{t*}_{nj}$, and $\mathbf{S}^{r*}_{nj}\mathbf{U}^F_{nj}$ respectively in Problem~\ref{pro:inter} and~\ref{pro:intra}, where $\{\mathbf{V}^{C*}_{nj}\}$, $\{\mathbf{S}^{t*}_{nj}\}$, and $\{\mathbf{S}^{r*}_{nj}\}$ are the solutions of Problem~\ref{pro:userspace}.

\begin{Thm}\emph{(Connetction of the Original Problem and the Subproblems in Partially Connected Networks)}
For partially connected MIMO cellular networks, we have with probability 1, the solutions of Subproblem~\ref{pro:userspace},~\ref{pro:inter},~\ref{pro:intra}, i.e. $\{d^*_{gk}\}$, $\{\mathbf{U}^*_{gk}\}$, $\{\mathbf{V}^*_{gk}\}$, are also valid solution of Problem~\ref{pro:original}. Hence, the performance of the decomposed problems, i.e. $\sum_{g=1}^{G}\sum_{k=1}^Kd^*_{gk}$ gives a lower bound of that of the original problem.\label{thm:equal2}
\end{Thm}
\proof Please refer to Appendix\ref{pf_thm:equal2}.
\endproof

\subsection{Solution to Problem~\ref{pro:userspace} (Stream Assignment and Subspaces Design Problem)}
We extend the greedy-based Algorithm~\ref{alg:offline_f} to cater for the partial connectivity in Problem~\ref{pro:userspace}.

\begin{Alg}[Greedy-Based Solution for Problem~\ref{pro:userspace}]
\label{alg:offline_p}
\begin{itemize}
\item[]
\item{\bf Step 1 Initialization:} Initialize the
number of streams as: $d_{nj}=\min(\mbox{rank}(\mathbf{H}_{nj,n}),d^{\max}_{nj})$, $\forall n\in\{1,...,G\}$, $j\in\{1,...,K\}$.
\item{\bf Step 2 Calculate the common null spaces:}
 At each BS $n\in\{1,...,G\}$, calculate the intersection of the null spaces of the inter-cell cross links,
 i.e. $\mathcal{N}_n(\mathbb{M})=\cap_{
(g,k)\in\mathbb{M}}\mathcal{N}^t_{gk,n}$, $
\mathbb{M} \subseteq \{(g,k):\;g\neq n\in\{1,...,G\},
k\in\{1,...,K\}\}$, as follows:
\begin{itemize}
\item Denote $\mathbb{M}_{n}=\{(g,k):
\mathbf{H}_{nm}\neq \mathbf{0}\}$. Initialize
$\mathcal{N}_n(\emptyset)=\mathbb{C}^{N^t_n}$,
$\mathcal{N}_n(\{(g,k)\})=\mathcal{N}^t_{gk,n}$, and set the cardinality parameter $C=2$.
\item For every $\mathbb{M}\subseteq\mathbb{M}_{n}$ with
$|\mathbb{M}|=C$, if all the subsets of $\mathbb{M}$
with cardinality $(C-1)$ are not $\{0\}$, calculate
$\mathcal{N}_n(\mathbb{M})=\mathcal{N}_n(\mathbb{M}\backslash\{(g',k')\})\cap
\mathcal{N}(\{(g',k')\})$, where $(g',k')$ is an arbitrary element
in $\mathbb{K}_{sub}$. Update $C=C+1$. Repeat this process until
$\mathcal{N}(\mathbb{M})=\{0\}$,
$\forall\mathbb{M}\subseteq\mathbb{M}_{n}$ with
$|\mathbb{M}|=C$ or $C=|\mathbb{M}_{n}|$.
\item For every $\mathbb{M}\subseteq\mathbb{M}_{n}$ with
$\mathcal{N}_n(\mathbb{M})\neq \{0\}$,
set
$\mathcal{N}_n(\mathbb{M}\cup(\{1,...,K\}\backslash\mathbb{M}_{n}))
=\mathcal{N}_n(\mathbb{M})$.
\end{itemize}

At each MS $M_{gk}$, calculate $\mathcal{N}^r_{gk}(\mathbb{M}')=\cap_{
n\in\mathbb{M}'}\mathcal{N}^r_{gk,n}$, $
\mathbb{M}' \subseteq \{n:\;n\neq g\in\{1,...,G\}\}$ using a similar process.
\item{{\bf Step 3 Design  $\mathbb{V}^C_{nj}$ (}i.e. $\mbox{span}(\mathbf{V}^C_{nj})${\bf) :}}
At BS $n$, $n\in\{1,...,G\}$,  design
$\mathbb{V}^C_{nj}$, $j\in\{1,...,K\}$ one by one as follows: For the $j$-th MS of BS-$n$,
\begin{itemize}
\item Update the number of streams assigned to the $j$-th MS of BS-$n$ if there is not enough signal dimension
left, i.e. update
$d_{nj}=\min\left(d_{nj},N^t_g-\mbox{dim}\left((+_{k<j}\mathbb{V}^C_{nk})
+\mathcal{N}(\mathbf{H}_{nj})\right)\right)$;
\item Design $\mathbb{V}^C_{nj}$ based on the principles
that {\bf A)} $\mathbb{V}^C_{nj}$ is orthogonal to the previous
designed core spaces and is contained by the effective subspace of
the direct link, i.e.
$\mathbb{V}^C_{nj}\subseteq\left((+_{k<j}\mathbb{V}^C_{nk})+\mathcal{N}(\mathbf{H}_{nj})\right)^\perp$;
{\bf B)} A subspace which belongs to a null space $\mathcal{N}(\mathbb{M})$
with larger ``weight" (i.e. $W_n(\mathcal{N}(\mathbb{M}))$, defined below)
is
selected with higher priority.
\begin{eqnarray}
W_n(\mathcal{N}(\mathbb{M}))=\sum_{(g,k)\in\mathbb{M}}\min\left(d_{gk},\mbox{rank}(\mathbf{H}_{gk,n})\right)
\label{eqn:weight}
\end{eqnarray}

From the left hand side of
\eqref{eqn:feasible_p}, this weight is the maximum number of IA
constraints that one can mitigate by selecting a one dimensional
subspace in
$\mathcal{N}(\mathbb{M})$.
 \end{itemize}

\item{\bf Step 4 Design $\mathbb{S}^t_{nj}$ and $\mathbb{S}^r_{gk}$ (}i.e. $\mbox{span}(\mathbf{S}^t_{nj})$, $\mbox{span}(\mathbf{S}^r_{gk})${\bf):}

At BS $n$, $n\in\{1,...,G\}$, design $\{\mathbb{S}^t_{nj}\}$:
\begin{itemize}
\item {\bf{A.} } Generate a series of potential
$\mathbb{S}^t_{nj}(d),
d\in\{0,1,...,N^t_g-\sum_{k=1}^Kd_{nk}\}$ with
$\mbox{dim}\left(\mathbb{S}^t_{nj}(d)\right)=d$ based on the
principles that {\bf A)} $\mathbb{S}^t_{nj}\subseteq\left((+_{k\in\{1,...,K\}}\mathbb{V}^C_{nk})\right)^\perp$, {\bf B)} Same as the principle B in Step 3.

\item {\bf{B.} } \emph{Choose the best $\mathbb{S}^t_{nj}$:} Set $\mathbb{S}^t_{nj}=\mathbb{S}^t_{nj}(d^*)$,
 where
 \begin{eqnarray}\nonumber d^*&=&\arg\max_{d} \left(d_{nj}
 d-\frac{}{}\right.\sum_{g=1,\neq
n}^G\sum_{k=1}^K\min\left(d_{gk},\mbox{rank}(\mathbf{H}_{gk,n})\right)
 \\&&\left.\times\min\left(d_{nj}, \left|(\mathbb{V}^C_{nj}+
\mathbb{S}^t_{nj}(d))\cap(\mathcal{N}^t_{gk,n})^\bot\right|\right)\frac{}{}\right).
\label{eqn:d_t}\end{eqnarray}
 \end{itemize}

Similarly, at each MS $M_{gk}$, generate $\mathbb{S}^r_{nj}(d),
d\in\{0,1,...,N^r_{gk}-d_{gk}\}$ based on principle B. Set $\mathbb{S}^r_{gk}=\mathbb{S}^r_{gk}(d^*)$,
 where
 \begin{eqnarray}\nonumber d^*&=&\arg\max_{d} \left(d_{gk}
 d-\frac{}{}\right.\sum_{n=1,\neq
g}^G\sum_{j=1}^K\min\left(d_{gk},\left|\mathbb{S}^r_{gk}\cap(\mathcal{N}^r_{gk,n})^\bot\right|\right)
 \\&&\left.\times\min\left(d_{nj}, \left|(\mathbb{V}^C_{nj}+
\mathbb{S}^t_{nj}(d))\cap(\mathcal{N}^t_{gk,n})^\bot\right|\right)\frac{}{}\right).
\label{eqn:d_r}\end{eqnarray}

\item{\bf Step 5 IA Feasibility checking:}
Similar to Step 3 in Algorithm~\ref{alg:offline_f}, set $v^t_{nj}=d_{nj}S^t_{nj}$,
$v^r_{gk}=d_{gk}S^r_{gk}$, where $S^t_{nj}$ and  $S^r_{gk}$ are defined in Definition~\ref{def:trans_partial}. Set $c_{gk,nj}=\min\left(d_{gk},\left|\mbox{span}(\mathbf{S}^r_{gk})\right.\right.$\\ $\left.\left.\cap
(\mathcal{N}^r_{gk,n})^\bot\right|\right)\min\left(d_{nj},
|(\mbox{span}(\mathbf{V}^C_{nj})+\mbox{span}(\mathbf{S}^t_{nj}))\cap(\mathcal{N}^t_{gk,n})^\bot|
\right)$, if $g\neq n$; $c_{gk,nj}=0$, otherwise. Use the low complexity algorithm in Appendix\ref{alg_detail} to check if the system is IA feasible.  If the network is not feasible, go to
Step 6. Otherwise, set $d_{nj}^*=d_{nj}$, and set
$\mathbf{V}^{C*}_{nj}$, $\mathbf{S}^{t*}_{nj}$, $\mathbf{S}^{r*}_{nj}$ to be matrices
aggregated by the basis vectors of $\mathbb{V}^{C}_{nj}$, and
$\mathbb{S}^t_{nj}$, $\mathbb{S}^r_{nj}$, respectively, $\forall
n\in\{1,...,G\},j\in\{1,...,K\}$. Exit the algorithm.

\item{\bf Step 6 Update stream assignment:}
Update $d_{g'k'}=d_{g'k'}-1$ and go back to Step
2, where $(g',k')$ is given by (the first line of) \eqref{eqn:selectn}.
\end{itemize}
\end{Alg}

\begin{Remark}[Subspace Design Criterion in Algorithm~\ref{alg:offline_p}]
Similar to the stream assignment criteria \eqref{eqn:selectn}, the core space $\{\mathbf{V}^{C}_{nj}\}$ and free space $\{\mathbf{S}^t_{nj}\}$ should be designed to alleviate the IA
feasibility constraint as much as possible  in order to enhance the network DoF. Hence, both \eqref{eqn:weight} and \eqref{eqn:d_t} are designed to maximize the
 difference between the number of freedoms in intermediate precoder design
 minus the number of inter-cell IA constraints.\hfill~\IEEEQED
\end{Remark}

\begin{Remark}[Relationship between Algorithm~\ref{alg:offline_f} and~\ref{alg:offline_p}]
\label{rem:relation} In fact, Algorithm~\ref{alg:offline_p} is a
backward compatible extension of Algorithm~\ref{alg:offline_f}. When
the network is fully connected, Step  2$\sim$4 in
Algorithm~\ref{alg:offline_p} will generate $\{\mathbf{V}^C_{nj}\}$,
$\{\mathbf{S}^t_{nj}\}$, and $\{\mathbf{S}^r_{nj}\}$ with $\mathbf{S}^t_{nj}=\mathbf{S}_n$, $\forall
j\in\{1,...,K\}\}$ and $\mbox{rank}(\mathbf{S}^r_{nj})=N^r_{nj}$. However, this particular choice of the core
space will not offer any additional DoF gain compared to other choices of $\{\mathbf{V}^I_{nj}\}$ and
 $\{\mathbf{S}_{n}\}$ satisfying constraint
 \eqref{eqn:orth}  in the fully connected case.\hfill~\IEEEQED
\end{Remark}

\subsection{Solution of Subproblems~\ref{pro:inter} and \ref{pro:intra} in Partially Connected Networks}
The solution to Problems~\ref{pro:inter} and \ref{pro:intra} in the partially connected networks are very similar to Algorithm 3 and 4, respectively. Details are omitted to avoid redundance.

\section{Performance Analysis}
\subsection{Symmetric MIMO Cellular Networks with Partial Connectivity}
\begin{Def} [Symmetric MIMO Cellular Networks with Partial Connectivity]
\label{def:sym} A symmetric MIMO cellular network has $G$ BSs (each with $N^t$ antennas) serving $K$ MS (each with $N^r$ ($\le N^t$) antennas) per BS. There are at most $d^{\max}_{gk}=d_f$ data streams per MS.
The partial connectivity is induced by the {\em path loss} effects as well as the {\em transmit spatial correlation} effects due to local scattering\footnote{The transmit spatial correlation is caused by the lack of scattering in the propagation environment around the BSs.} \cite{Local3,Ruan2}.
Links from BS-$n$ to MSs of BS-$g$ with $J<|n-g|<G-J$ are assumed to have huge path losses and they are regarded as not connected. Furthermore, $R_1(\le N^r)$ and $R_2(\le N^r)$ denote the ranks of the intra-cell links and inter-cell links.  ~\hfill \IEEEQED
\end{Def}

As a result, there are three key parameters, i.e. $J$, $R_1$ and $R_2$, which characterize the connection density, the rank of the intra-cell and inter-cell links of the symmetric MIMO cellular network, respectively. In particular, the BS side partial connectivity in Definition~\ref{def:model} of the above symmetric network is given by:
\begin{eqnarray}
\mathcal{N}^t_{gk,n}=
\left\{\begin{array}{l}\mbox{span}\left(\{\mathbf{n}(q):
q\in\mathbb{R}_1(k)\} \right),
 \mbox{ if: } g=n
\\ \mbox{span}\left(\{\mathbf{n}(q):q\in\mathbb{R}_2(n-g)\}
\right), \mbox{ if: } g\neq n, 0<|n-g|\le J \mbox{ or }|n-g|\ge G-J
\\\mathbb{C}^{N^t}, \mbox{
otherwise.}\end{array}\right.\label{eqn:NT_a}
\end{eqnarray}
where $\{\mathbf{n}(q)\}$,
$q\in\{0,1,...,N^t-1\}$ form a basis for $\mathbb{C}^{N^t}$,
\begin{eqnarray}\nonumber \mathbb{R}_1(k)=\{0,1,...,N^t-1\}\backslash\{kR_1\mbox{ mod } N^t,kR_1+1\mbox{ mod } N^t,...,(k+1)R_1-1\mbox{ mod } N^t \},
\\\nonumber \mathbb{R}_2(m)=\{0,1,...,N^t-1\}\backslash\{mR_2\mbox{ mod } N^t,mR_2+1\mbox{ mod } N^t,...,(m+1)R_2-1\mbox{ mod } N^t \},\end{eqnarray}
$\mbox{span}(\{\mathbf{n}\})$ denotes the linear space
spanned by the vectors in set $\{\mathbf{n}\}$. To make sure the direct links can have sufficient rank, we also assume:
 and $d_fK\le N^t$.

\begin{Remark}[Partial Connectivity in Practice] In practice, singular values of channel matrices or the path gain of links can hardly be \emph{absolutely} 0, and hence, the DoF defined by the asymptotic slope of the throughput-SNR curve may not correspond to the number of data streams transmitted. However, this shall not jeopardize the value of the proposed algorithm, i.e. Algorithm~\ref{alg:inter_full},~\ref{alg:intra_full}, and~\ref{alg:offline_p}. This is because in practice, we are interested in the performance at \emph{finite} SNR regime only. As long as the singular values or the path gains are below a sufficiently small threshold, we shall quantize the singular values and the path gain to be zero and the said channel is partially connected according to Definition~\ref{def:model}.~\hfill~\IEEEQED
\end{Remark}

\subsection{Analytical Results}
\begin{Thm}[Achievable DoF of the Proposed Scheme] \label{thm:DoF}
The total DoF achieved by the proposed
scheme in a symmetric MIMO cellular network in Definition~\ref{def:sym} is lower bounded by $GKd^*$, where $d^*$ is the number of streams assigned to each MS, given by:
\begin{eqnarray}
d^*=\min\left(R_1,\left\lfloor\max\left(\frac{N^r}{\min(G-1,2J) K \frac{R_2}{N^t}+1},\frac{N^r+N^t}{\min(G-1,2J) K+2}
\right)\right\rfloor\right). \label{eqn:result1}
\end{eqnarray}
\end{Thm}
\proof Please refer to Appendix\ref{pf_thm:DoF} for the proof.
\endproof
The following are some interpretations of the results in  \eqref{eqn:result1}.
\begin{Remark}[Backward Compatibility with Fully Connected K-pair Interference Channels]
Consider a special case of fully connected $G$-pair interference channel with $K=1$, $J\ge\frac{G}{2}$,
$R_1=R_2=N^r$. The achievable DoF in  \eqref{eqn:result1} reduces to $\left\lfloor\frac{N^t+N^r}{G+1}
\right\rfloor$, which is consistent with result in the conventional IA feasibility condition\footnote{Using the conventional IA feasibility condition in \cite{JafarDf} for $G$-pair MIMO interference channels, we have $N^t+N^r-(G+1)d\ge 0 \Rightarrow d\le\frac{N^t+N^r}{G+1}$.}.
~\hfill \IEEEQED
\end{Remark}

\begin{Remark}[How Partial Connectivity Affects DoF]
\label{remark:corhelp} When the partial connectivity effect is strong, i.e. $J< \frac{G}{2}$, $R_2\ll N^t$, the network total DoF becomes
$GK\min\left(R_1,\left\lfloor\frac{N^r}{2JK \frac{R_2}{N^t}+1}\right\rfloor\right)$.
Hence, it can be observed that partial connectivity affects
the total DoF in three aspects:
\begin{itemize}
\item\emph{Gain due to the
connection density}: As the connection density parameter $J$ limits the
maximum number of MSs that each BS may interfere, the total DoF of the MIMO cellular network is $\mathcal{O}(G)$, which scales with the number of the BS. This behavior represents a significant gain  compared with the fully connected case in which the total DoF$= \mathcal{O}(1)$ \cite{JafarDf}.
\item\emph{Gain due to weak
inter-cell links}: When the network is dense, i.e.
$J\gg 1$, $\frac{N^r}{2LK \frac{R_2}{N^t}+1}\simeq \frac{N^rN^t}{2LKR_2}$. Hence, a $\frac{N^t}{R_2}$ factor gain
can be further observed.
\item\emph{Loss due to weak intra-cell links:} Note that the rank of the direct link $R_1$ is one of the terms in $\min$ function and hence, the partial connectivity may also limit the
system DoF when the intra-cell links are weak, i.e. small $R_1$.
 ~\hfill \IEEEQED
\end{itemize}
\end{Remark}
{\color{mblue}
\begin{Remark}[DoF Scaling Law w.r.t. Number of Antennas]
\begin{itemize}
\item[]
\item{\em Strong inter-cell link case:} When the inter-cell links are strong, i.e. $R_2\simeq N^t$, in \eqref{eqn:result1}, the second term is the $\max$ operation is larger, hence, the total DoF scales on $\mathcal{O}(N^r+N^t)$.
\item{\em Weak inter-cell link case:} When the inter-cell links are weak, i.e. $R_2\ll N^t$, in \eqref{eqn:result1}, the first term is the $\max$ operation is larger, hence, the total DoF scales on $\mathcal{O}(N^rN^t)$.
\end{itemize}
 Comparing the two cases, we can see that antennas are more ``effective" when the inter-cell links are weak. This is because when inter-cell links are weak, the partial connectivity can be exploited to eliminate part of the potential interference, thus alleviating the constraints on transceiver design.
 ~\hfill \IEEEQED
\end{Remark}
}

\section{Numerical Results}
\begin{figure} \centering
\includegraphics[scale=0.7]{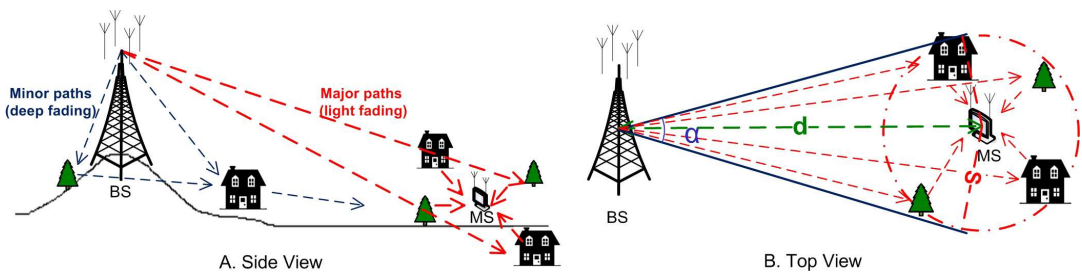}
\caption {\emph{Local scattering} effect: {The
lack of scattering in the propagation environment leads to spatial channel
correlation.}} \label{fig_channel1}
\end{figure}

In this section, we shall illustrate the performance of the proposed
scheme by simulation.
\begin{Def}[Randomized Partially Connected MIMO Interference Channels]
\label{def:model2} Consider a MIMO cellular
network with $G$ BSs and $GK$ MSs. Each BS has
$N^t$ antennas and each MS has $N^r$
antennas and requests $d_f$ data streams.
The BSs and MSs are distributed uniformly in a $30km\times30km$
area. All BSs transmit at power $P$. Denote $D_{gk,n}$ as the
distance between the BS $n$ and the $k$-th MS of BS-$g$. The network is partially connected due to:
\begin{itemize}
\item {\bf Path loss effect:}
If $D_{gk,n} > L$, we assume the channel from the BS $n$ to the $k$-th MS of BS-$g$
is not connected, i.e. $\mathbf{H}_{gk,n} =
\mathbf{0}$.
\item {\bf Local scattering effect:}
If $D_{gk,n} \le L$, due to local scattering effect, channel fading are correlated (only transmit correlation), and hence:  \begin{eqnarray}\mathcal{N}^t_{gk,n}=\mathbb{L}_{gk,n},\;\mathcal{N}^r_{gk,n}=\{0\}
\end{eqnarray}
where $\mathbb{L}_{gk,n}=\mbox{span}\left(\{\mathbf{e}_{N^t}(q): q\in\{1,...,N^t\}\mbox{ and satisfies } \eqref{eqn:virtual_1}\}\right)$, $\mathbf{e}_N(\omega)=\frac{1}{\sqrt{N}}[1,$ $e^{-j2\pi(\omega)}$ $,
e^{-j2\pi(2\omega)}...,e^{-j2\pi((N-1)\omega)}]^t$.
\begin{eqnarray}
\frac{1}{N^t}<\left\lfloor\left|\frac{\sin\theta}{2}-\frac{q}{N^t}\right|\right\rfloor<\frac{N^t-1}{N^t}
,\;\forall \theta\in [\theta_{gk,n}-F_a(S,d_{gk,n}),
\theta_{gk,n}+F_a(S,d_{gk,n})]
   \label{eqn:virtual_1}
\end{eqnarray}
where $F_a(S,d_{gk,n})=\left\{\begin{array}{l}
 \arcsin\frac{S}{d_{gk,n}} \mbox{ when: }S\le d_{gk,n}
 \\\pi \;\;\;\; \;\;\;\; \;\;\;\;\;\;\mbox{ when: }S> d_{gk,n}\end{array}\right.$, $S$ is the \emph{local scattering radius} as illustrated in Fig.~\ref{fig_channel1}B, $\theta_{gk,n}$ is the angle between the antenna array normal
direction and the direction from BS $n$ to the $k$-th MS of BS-$g$. Please refer to \cite{Ruan2} for details.
~\hfill \IEEEQED
\end{itemize}
\end{Def}

The proposed scheme is compared with 5
reference baselines below{\color{mblue}\footnote{\color{mblue}The feasible bound for IA algorithms on partially connected MIMO cellular network is still unknown. Therefore, we cannot plot the theoretical upper bound as one of the benchmarks in simulation.}$^,$\footnote{\color{mblue} Note that BL 1 is a simplification of the proposed scheme. It does not address the partial connectivity issue. BL 2 is generalized from \cite{JafarD1} to the MIMO Cellular Network. Comparison with BL1 illustrates the importance of exploiting partial connectivity. On the other hand, comparison with BL2 illustrates the necessity of the decomposition approach proposed in this paper.}}:
\begin{itemize}
\item{\bf Simplified proposed scheme (Baseline (BL) 1):}
The stream assignment and transceiver matrices are designed by Algorithms~\ref{alg:offline_f}, \ref{alg:inter_full} and \ref{alg:intra_full}. As we illustrate in Remark~\ref{rem:relation}, Algorithm~\ref{alg:offline_f} is a simplified version of Algorithm~\ref{alg:offline_p}.
\item{\bf Naive extension of the existing IA algorithm \cite{JafarD1} (BL 2):} The transceivers are designed by naive extension of iterative IA algorithm in \cite{JafarD1} as described in Algorithm~\ref{pro:naive}.
\item{\bf Coordinated beamforming \cite{J_CorBeam} (BL 3):} The BSs jointly optimize their precoders to improve
the overall system SINR performance using the algorithm proposed in \cite{J_CorBeam}.
\item {\bf Round robin scheduling with Intra-cell zero-forcing (BL 4):} The BSs are scheduled to transmit using round robin. Zero-forcing precoders are adopted.
\item{\bf Isotropic transmission (BL 5):}  The BSs and the MSs apply random precoders and decorrelators, respectively.
\end{itemize}
\subsection{Fully Connected MIMO Cellular Network}
Fig.~\ref{fig_fullyconnected}
illustrates the sum throughput versus SNR ($10\log_{10}(P)$) for the
proposed scheme and 5 baselines for an IA feasible MIMO cellular network with $G=3$, $K=2$, $d_f=1$, $N^t=5$, $N^r=2$, $L,S\ge 30\sqrt{2}
km$. BL 4 can only achieve 2 DoF as each BS has only
$\frac{1}{3}$ of the time to transmit. BL 2 achieves only 3 DoF due to the cross link - direct link
overlapping issue. The throughput of BL 3 saturates at high SNR since coordinated beamforming \cite{J_CorBeam} can only suppress part of the interference. {\color{mblue}On the other hand, the proposed algorithm and BL 1 achieve 6 DoF, which is an achievable upper bound.} This result also confirms the comments made in
Remark~\ref{rem:relation} that Algorithm~\ref{alg:offline_p} is a backward compatible extension of
Algorithm~\ref{alg:offline_f} and they have the same performance in fully connected networks.
\subsection{Partially Connected MIMO Cellular Network}
\subsubsection{\color{mblue}Performance w.r.t. SNR}
\begin{figure} \centering
\includegraphics[scale=0.63]{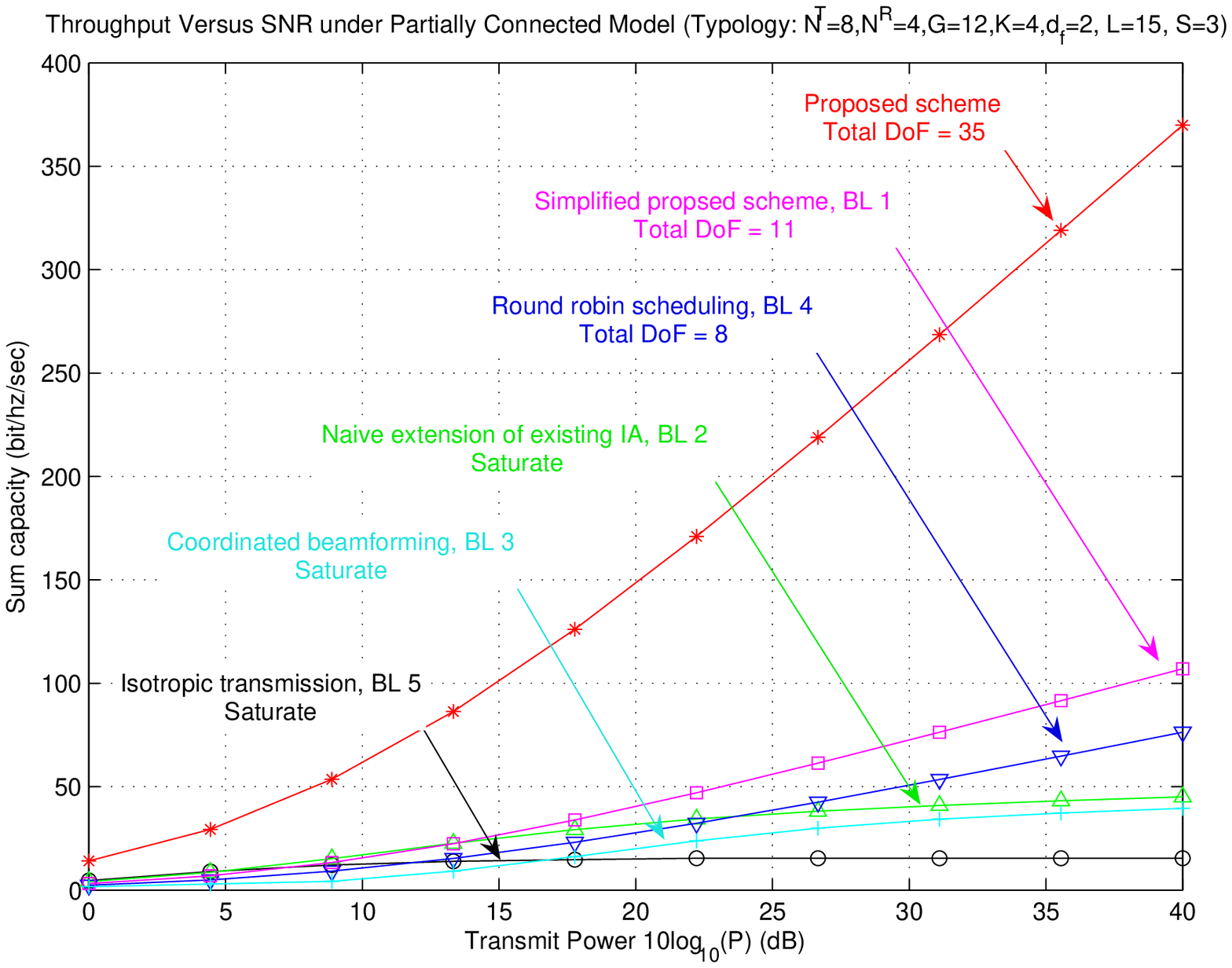}
\caption {Sum throughput versus SNR for the proposed algorithm (and
5 baselines)  in a partially connected MIMO cellular network.} \label{fig_partial}
\end{figure}

\begin{figure} \centering
\includegraphics[scale=0.6]{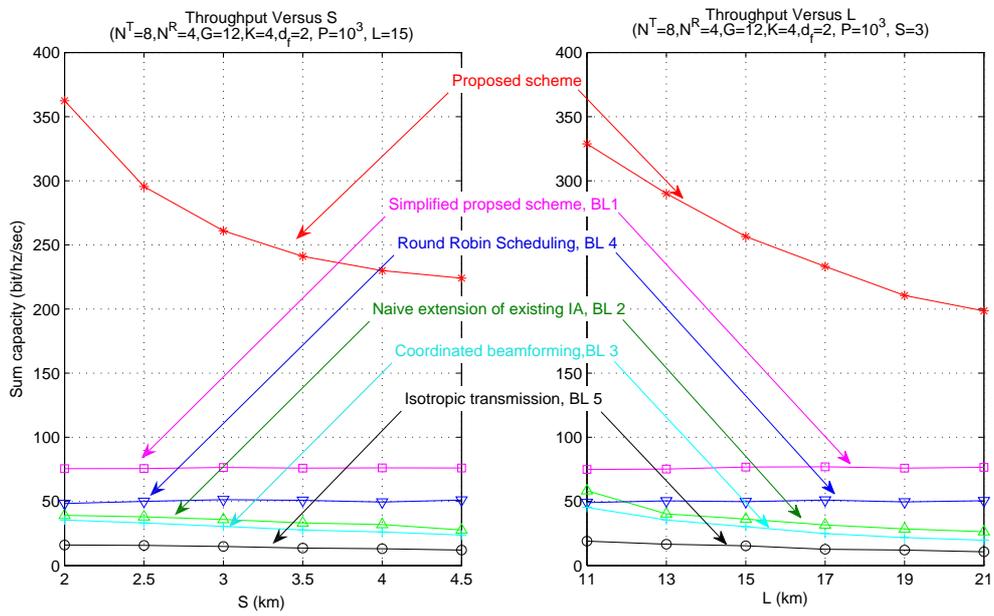}
\caption {\color{mblue}Sum throughput versus partial connectivity parameters for the proposed algorithm (and
5 baselines). $L$ represents the connection density and $S$ is local scattering radius (spatial correlation decreases when $S$ increases). } \label{fig_SL}
\end{figure}

Fig.~\ref{fig_partial} illustrates the sum throughput versus SNR  ($10\log_{10}(P)$) for the
proposed scheme and 5 baselines  in a MIMO cellular network with $G=12$, $K=4$, $d_f=2$, $N^t=8$, $N^r=4$, $L=15 km$, $S=3 km$. The throughput of BL 2 also saturates at high SNR since the network is not IA feasible. BL 1 achieves 11 DoF only as Algorithm~\ref{alg:offline_f} fails to exploit the benefit of partial connectivity. {\color{mblue}On the other hand, the proposed algorithm achieves 35 DoF, which is significantly higher than all the baselines. Furthermore, the total DoF upper bound for the fully connected MIMO network is only 11. This demonstrates that partial connectivity can indeed contribute to the significant gain in system throughput.} The comparison between the proposed scheme and BL 1 illustrates the importance of incorporating partial connectivity topology in the IA algorithm.
{\color{mblue}\subsubsection{Performance w.r.t. Partial Connectivity Factors}
To better illustrate how different partial connectivity factors such
as path loss and spatial correlation affect system performance, we
illustrate the sum throughput versus $L$ (the maximum distance that
a BS can interfere a MS) and $S$ (the radius of the local
scattering) under a fixed SNR (30dB) in Fig.~\ref{fig_SL}. By comparing the performance of
the proposed scheme with different partial connectivity parameters,
we have that the performance of the proposed scheme roughly scales
$\mathcal{O}\left(\frac{1}{LS}\right)$, which illustrates a
consistent observation as Remark~\ref{remark:corhelp} that weaker
connectivity can indeed contribute to higher system
performance. Moreover, comparison of the proposed algorithm with
BL 1 further illustrates the importance of adapting the transceiver structures given in Def.~\ref{def:trans_partial} to exploit partial connectivity. By dynamically adapting the transceiver structures, the proposed scheme
obtains significant performance gain over a wide range of partial
connectivity levels.}

\appendices
 \section*{Appendices}
\subsection{Proof for Theorem~\ref{thm:equal}}
\label{pf_thm:equal}
\begin{Lem2}[IA Feasibility Conditions of MIMO Cellular Network]\label{lem:c2p_f} With i.i.d. fading, {\color{mblue}Problem~\ref{pro:original}} is equivalent to the following problem almost surely.
\begin{Prob2}[Transformed IA Problem]\label{pro:p2p_f}
\begin{eqnarray}
&&\max_{\{d_{nj}\},{\color{mblue}\{\mathbf{V}'_{nj}\}},\{\mathbf{U}'_{gk}\}}\sum_{n=1}^{G}\sum_{j=1}^{K}d_{nj}
\\&\mbox{s.t.:}&\mbox{rank}(\mathbf{U}'_{gk})=d_{gk},\;{\color{mblue}\mbox{rank}([\mathbf{V}'_{n1},...
\mathbf{V}'_{nK}])=\sum_{j=1}^Kd_{nj}}\label{eqn:rank_p2p}\\
&&(\mathbf{U}'_{gk})^H\mathbf{H}_{gk,n}\mathbf{V}'_{nj}=\mathbf{0}\label{eqn:cross_p2p}
\end{eqnarray}
$d_{nj}\in\{0,1,...,d^{\max}_{nj}\}$, $\forall n,g\in\{1,...,G\}, n\neq g,  k,j\in\{1,...,K\}$.\hfill~\IEEEQED \end{Prob2}
\end{Lem2}
{\color{mblue}\proof We need to show that {\bf a)} if $\{d_{nj},\mathbf{U}_{gk},\mathbf{V}_{nj}\}$ is a solution of Problem~\ref{pro:original}, there must exists $\{d_{nj},\mathbf{U}'_{gk},\mathbf{V}'_{nj}\}$ which is a solution of Problem\ref{pro:p2p_f}, and {\bf b)} vise versa.
\begin{itemize}
\item {Proof of \bf a)}: Denote the transceivers in the solution of Problem~\ref{pro:original} as $\{\mathbf{U}^*_{nj}, \mathbf{V}^*_{nj}\}$. Let $\mathbf{U}'_{nj}=\mathbf{U}^*_{nj}$, $\mathbf{V}'_{nj}=\mathbf{V}^*_{nj}$, then from \eqref{eqn:rank_1}, \eqref{eqn:cs_cross_1} we have \eqref{eqn:rank_p2p} and \eqref{eqn:cross_p2p}.
\item {Proof of \bf b)}:  Denote the solution of Problem\ref{pro:p2p_f} as  $\{d^*_{nj}, \mathbf{U}'^*_{nj}, \mathbf{V}'^*_{n}\}$. Note that $\{\mathbf{U}'^*_{nj}, \mathbf{V}'^*_{n}\}$ are functions of the cross link channel states, i.e. $\{\mathbf{H}_{gk,n}:g\neq n\}$, which are independent of the direct link channel states, i.e. $\{\mathbf{H}_{nk,n}\}$. Hence we have
\begin{eqnarray}
\mbox{rank}\left(\left[\begin{array}{c}(\mathbf{U}'^*_{n1})^H\mathbf{H}_{n1,n}
\\\dotfill
\\(\mathbf{U}'^*_{nK})^H\mathbf{H}_{nK,n}\end{array}\right]
[\mathbf{V}'^{*}_{n1},...\mathbf{V}'^{*}_{nK}]\right)
&=&\sum_{j=1}^Kd_{nj}, \;\forall n\in\{1,..,G\}
\label{eqn:sufficient_rank_p2p}
\end{eqnarray}
almost surely. Let $\mathbf{U}^*_{nj}=\mathbf{U}'^*_{nj}$, $\mathbf{V}^{I*}_{nj}=\mathbf{V}'^*_{nj}$ and use Algorithm~\ref{alg:intra_full} to construct $\mathbf{V}^*_{nj}$. Then from \eqref{eqn:sufficient_rank_p2p}, Theorem~\ref{thm:intra_full} holds, which means $\{\mathbf{U}^*_{nj}, \mathbf{V}^*_{nj}\}$ satisfies \eqref{eqn:viso}, \eqref{eqn:uvrank2}, \eqref{eqn:spaceeq} and \begin{eqnarray}
(\mathbf{U}^*_{nj})^H\mathbf{H}_{nj,n}\mathbf{V}_{nk}=\mathbf{0}, \;\forall n\in\{1,...,G\},\; \forall j\neq k\in\{1,...,K\}.\label{eqn:intra_A}
\end{eqnarray}

From \eqref{eqn:spaceeq}, \eqref{eqn:cross_p2p} we have:
\begin{eqnarray}
(\mathbf{U}^*_{gk})^H\mathbf{H}_{gk,n}\mathbf{V}_{nj}=\mathbf{0}, \;\forall n\neq g\in\{1,...,G\},\; \forall j,k\in\{1,...,K\}.\label{eqn:inter_A}
\end{eqnarray}

From \eqref{eqn:viso}, \eqref{eqn:uvrank2}, \eqref{eqn:intra_A}, and \eqref{eqn:inter_A}, we have that $\{\mathbf{U}^*_{nj}, \mathbf{V}^*_{nj}\}$ satisfy \eqref{eqn:rank_1}$\sim$\eqref{eqn:normv}.\endproof
\end{itemize}}
From Lemma\ref{lem:c2p_f}, we need to show {\bf A)} an optimizing solution $\{d^*_{gk},\mathbf{U}'^*_{g},\mathbf{V}'^*_{gk}\}$, of Problem\ref{pro:p2p_f} is a feasible solution of the Subproblems~\ref{pro:user},~\ref{pro:inter},~\ref{pro:intra} and {\bf B)} vise versa.

First consider the statement A). We first have two lemmas:
\begin{Lem2}[Non-overlapped Subspaces]\label{lem:no_intersect} $\mathbb{V}$ is a subspace uniformly distributed in Grassmannian $\mathcal{G}(D,N)$. For any $N-D$ dimensional subspace $\mathbb{S}$, $\mathbb{V}\cap\mathbb{S}=\{0\}$ almost surely.
\end{Lem2}
\proof When $D\le \lfloor\frac{N}{2}\rfloor$, we have:
$\mathbb{V}\cap\mathbb{S}=\{0\}\Leftrightarrow \mathbf{v}\not\perp (\mathbb{S})^\perp,\;\forall \mathbf{v}\in\mathbb{V}
\Leftrightarrow \theta_{\max}(\mathbb{V},(\mathbb{S})^\perp)<\frac{\pi}{2}
$,
where $\theta_{\max}(\mathbb{A},\mathbb{B})$ denotes the largest principal angle between subspace $\mathbb{A}$ and $\mathbb{B}$. Note that both $\mathbb{V}$ and $(\mathbb{S})^\perp$ are $D$ dimensional subspaces with $D<\frac{N+1}{2}$. From Theorem 1 in \cite{PrincipleAngle}, we have $\Pr(\theta_{\max}(\mathbb{V},(\mathbb{S})^\perp)<\frac{\pi}{2})=1$.

Similarly, when $D> \lfloor\frac{N}{2}\rfloor$, we have:
$\mathbb{V}\cap\mathbb{S}=\{0\}\Leftrightarrow \mathbf{s}\not\perp (\mathbb{V})^\perp,\;\forall \mathbf{s}\in\mathbb{S}
\Leftrightarrow \theta_{\max}(\mathbb{S},(\mathbb{V})^\perp)<\frac{\pi}{2}$.
Note that both $\mathbb{S}$ and $(\mathbb{V})^\perp$ are $N-D$ dimensional subspaces with $N-D<\frac{N+1}{2}$ and $(\mathbb{V})^\perp$ uniformly distributes in Grassmannian $\mathcal{G}(N-D,N)$, we can again apply Theorem 1 in \cite{PrincipleAngle}. This completes the proof of the lemma.
\endproof
\begin{Lem2}[Uniformly Distributed Precoder Space]\label{lem:symmetric} Under the i.i.d. fading assumption, $\mbox{span}(\mathbf{V'}^*_{n})$, uniformly distributes in $\mathcal{G}(\sum_{j=1}^{K}d^*_{nj}, N^t_n)$, where $\mathbf{V'}^*_{n}$ and $\mathbf{d}^*_{nj}$ are the optimal solution of Problem\ref{pro:p2p_f}.
\end{Lem2}
\proof In Problem\ref{pro:p2p_f}, $\mbox{span}(\mathbf{V'}^*_{n})$ is a function of the channel states $\{\mathbf{H}_{gk,m}\}$, denote as $\mbox{span}(\mathbf{V'}^*_{n})=F(\{\mathbf{H}_{gk,m}\})$. For any two elements in $\mathcal{G}(\sum_{j=1}^{K}d^*_{nj}, N^t_n)$, $\mathbb{V}^a$ and $\mathbb{V}^b$, denote $\mathbb{H}_x=\{\{\mathbf{H}_{gk,m}\}: \mathbb{V}_x=F(\{\mathbf{H}_{gk,m}\})\}=\{\{\mathbf{H}_{gk,m}(x)\}\}$, where $x\in\{a,b\}$. Then there exists a unitary matrix $\mathbf{T}\in\mathbb{C}^{N^t_n\times N^t_n}$ such that
\begin{eqnarray}\mathbf{V}^b=\mathbf{T}\mathbf{V}^a, \mbox{ where } \mathbf{V}_x \mbox{ is the matrix aggregated by the basis of } \mathbb{V}_x, \;x\in\{a,b\}.\label{eqn:transit}
\end{eqnarray}

 Construct a mapping $G:\mathbb{H}^a \rightarrow \mathbb{H}^b, \mathbf{H}_{gk,m}(b)=\left\{\begin{array}{l}\mathbf{H}_{gk,m}(a)\mathbf{T} \mbox{, if: } m=n,\\\mathbf{H}_{gk,m}(a)\;\;\;\mbox{, otherwise. }\end{array}\right.$. Substitute \eqref{eqn:transit} into \eqref{eqn:cross_p2p}, we have that $G$ is bijective. Denote $D_h$ and $D_v$ as the probability density function of $\{\mathbf{H}_{gk,m}\}$ and $\mathbf{V'}^*_{n}$, respectively. Then from the i.i.d. fading assumption, we have that $D_h(G(\{\mathbf{H}_{gk,m}(a)\}))=D_h(\{\mathbf{H}_{gk,m}(a)\})$. Since $G$ is bijective, we have
 \begin{eqnarray}
 D_v(\mathbb{V}^b)=\int_{\mathbb{H}^b}D_h(x)\mbox{d}x=\int_{\mathbb{H}^a}D_h(G(x))\mbox{d}x=
 \int_{\mathbb{H}^a}D_h(x)\mbox{d}x=D_v(\mathbb{V}^a).\label{eqn:equal}
 \end{eqnarray}
 where $x$ denotes the elements in $\mathbb{H}^a$ or $\mathbb{H}^b$. With \eqref{eqn:equal}, we complete the proof.
\endproof

The constraints \eqref{eqn:rank_p2p}, \eqref{eqn:cross_p2p} in Problem\ref{pro:p2p_f} are the same as that addressed in \cite{JafarDf}, except that the number of transmitter and receiver are different when $K>1$. Yet, note that the analysis in \cite{JafarDf} can be easily extended to the case with different number of transmitters and receivers, Lemma\ref{lem:c2p_f} enables us to extend the existing IA feasibility conditions to the cellular case. Hence from Lemma\ref{lem:c2p_f} and the IA feasibility conditions obtained in \cite{JafarDf}, we have that the feasibility conditions of MIMO cellular network are given by \eqref{eqn:feasible1}.

Denote $\{d^*_{nj}\}$, $\{\mathbf{V'}^*_{n}\}$,
$\{\mathbf{U}'^*_{gk}\}$ as the optimizing variables for Problem\ref{pro:p2p_f}. Substitute the stream assignment policy $\{d^*_{nj}\}$ into Problem~\ref{pro:user}, the feasibility condition \eqref{eqn:feasible1} shall be satisfied almost surely. Consider the singular value decomposition of $\mathbf{V}'^*_{n}=\mathbf{A}^V_{n}\left[\begin{array}{c}\mathbf{S}^V_{n}\\\mathbf{0}
\end{array}\right](\mathbf{B}^V_{n})^H=
{\mathbf{A}'}^V_{n}\mathbf{S}^V_{n}
({\mathbf{B}}^V_{n})^H$, $\mathbf{U}'^*_{gk}=\mathbf{A}^U_{gk}\left[\begin{array}{c}\mathbf{S}^U_{gk}\\\mathbf{0}
\end{array}\right](\mathbf{B}^U_{gk})^H=
{\mathbf{A}'}^U_{gk}\mathbf{S}^U_{gk}
({\mathbf{B}}^U_{gk})^H$, where $\mathbf{A}^V_{n}$, $\mathbf{B}^V_{n}$, $\mathbf{A}^U_{gk}$, $\mathbf{B}^U_{gk}$ are $N^t_n\times N^t_n$, $\sum_{j=1}^{K}d^*_{nj}\times \sum_{j=1}^{K}d^*_{nj}$, $N^r_{gk}\times N^r_{gk}$, and $d^*_{gk}\times d^*_{gk}$ unitary matrices, respectively, ${\mathbf{A}'}^V_{n}$ and ${\mathbf{A}'}^U_{gk}$ are the first $\sum_{j=1}^{K}d^*_{nj}$ and $d^*_{gk}$ columns of the corresponding matrices, and $\mathbf{S}^V_{n}$ and $\mathbf{S}^U_{gk}$ are $\sum_{j=1}^{K}d^*_{nj}\times \sum_{j=1}^{K}d^*_{nj}$ and $d^*_{gk}\times d^*_{gk}$ diagonal matrices, respectively. Note that $\mbox{rank}(\mathbf{V}^*_{n})=\sum_{j=1}^{K}d^*_{nj}$, $\mbox{rank}(\mathbf{U}^*_{gk})=d^*_{gk}$, $\forall n,g,k$, we have $\mathbf{S}^V_{n}$ and $\mathbf{S}^U_{gk}$ are full rank. Hence we can set:
\begin{eqnarray}
\mathbf{V}_{n}&=&{\mathbf{A}'}^V_{n}=\mathbf{V}'^*_{n}{\mathbf{B}}^V_{n}(\mathbf{S}^V_{n})^{-1}
\label{eqn:vunitary}
\\\mathbf{U}_{gk}&=&{\mathbf{A}'}^U_{gk}
=\mathbf{U}'^*_{gk}{\mathbf{B}}^U_{gk}(\mathbf{S}^U_{gk})^{-1}\label{eqn:uunitary}
\end{eqnarray}

From Lemma\ref{lem:symmetric}, $\mbox{span}\left(\mathbf{V}'^*_{n}\right)$ uniformly distributes in $\mathcal{G}(\sum_{j=1}^K d^*_{nj},N^t_n)$. Note that $\mbox{span}\left(\mathbf{V}_{n}\right)
=\mbox{span}\left(\mathbf{V}'^*_{n}\right)$, $\mbox{dim}(\mbox{span}\left(\mathbf{S}_n\right))=N^t_n-\sum_{j=1}^K d^*_{nj}$, from Lemma~\ref{lem:no_intersect}, we have:
\begin{eqnarray}
\mbox{span}\left(\mathbf{V}_{n}\right)\cap\mbox{span}\left(\mathbf{S}_n\right)=\{0\}, \mbox{ almost surely.}\label{eqn:nocommon}
\end{eqnarray}

From equation \eqref{eqn:orth}, the columns of $\{\mathbf{V}^C_{nj}\}$ and $\mathbf{S}_n$ form a basis for $\mathbb{C}^{N^t_n}$. Hence, there exist matrices $\mathbf{R}_{n}$ ($\sum_{j=1}^K d^*_{nj}\times \sum_{j=1}^K d^*_{nj}$) and $\mathbf{Q}_n$ ($(N^t_n-\sum_{j=1}^K d^*_{nj})\times \sum_{j=1}^K d^*_{nj}$) such that:
\begin{eqnarray}
\mathbf{V}_{n}=
\left[\mathbf{V}^C_{n1},\mathbf{V}^C_{n2},...,\mathbf{V}^C_{nK},\mathbf{S}_n\right]
\left[\begin{array}{c}\mathbf{R}_n\\\mathbf{Q}_n\end{array}\right]=
\left[\mathbf{V}^C_{n1},\mathbf{V}^C_{n2},...,\mathbf{V}^C_{nK}\right]\mathbf{R}_n+\mathbf{S}_n\mathbf{Q}_n
\label{eqn:vaggreate}
\end{eqnarray}

From \eqref{eqn:nocommon}, $\mathbf{R}_n$ is full rank almost surely. Hence we have:
\begin{eqnarray}
\mathbf{V}_{n}\mathbf{R}^{-1}_n=
\left[\mathbf{V}^C_{n1},\mathbf{V}^C_{n2},...,\mathbf{V}^C_{nK}\right]+\mathbf{S}_n\mathbf{Q}_n\mathbf{R}^{-1}_n
\label{eqn:vf}
\end{eqnarray}

Set ${\mathbf{V}'}^F_{nj}=\mathbf{Q}_n\mathbf{R}^{-1}_{nj}$, where $\mathbf{R}^{-1}_{nj}$ is the matrix aggregated by the $(\sum_{k=1}^{j-1} d^*_{nj}) +1$ to $(\sum_{k=1}^{j} d^*_{nj})$-th column of $\mathbf{R}^{-1}$. Substitute $\{{\mathbf{V}'}^F_{nj}\}$,
$\{\mathbf{U}'_{gk}\}$ into Problem~\ref{pro:inter} and substitute $\{\mathbf{V}'_{nj}\}$, $\{\mathbf{U}'_{gk}\}$ into Problem~\ref{pro:intra}, From \eqref{eqn:rank_1}, \eqref{eqn:cs_cross_1}, \eqref{eqn:vunitary}, \eqref{eqn:uunitary} and \eqref{eqn:vf}, all the constraints are satisfied almost surely. This completes the proof for statement A).

Then to Statement B). From statement A) and \cite{JafarDf}, under the IA constraint \eqref{eqn:feasible1}, the optimal value of Problem~\ref{pro:inter}, \ref{pro:intra} are $0$ almost surely. Hence, denote $\{d^*_{nj}\}$, $\{\mathbf{V}^*_{nj}\}$,
$\{\mathbf{U}^*_{gk}\}$ as the corresponding outputs by solving the three problems sequentially.
Substitute these outputs to Problem~\ref{pro:original}, then from \eqref{eqn:uvrank2}, \eqref{eqn:intra}, and \eqref{eqn:inter_1}, we have that $\{d^*_{nj}\}$, $\{\mathbf{V}^*_{nj}\}$,
$\{\mathbf{U}^*_{gk}\}$ must satisfy \eqref{eqn:rank_1} and \eqref{eqn:cs_cross_1}. This completes the  proof.

\subsection{Low Complexity IA Feasibility Checking}
\label{alg_detail}
\begin{figure} \centering
\includegraphics[scale=0.75]{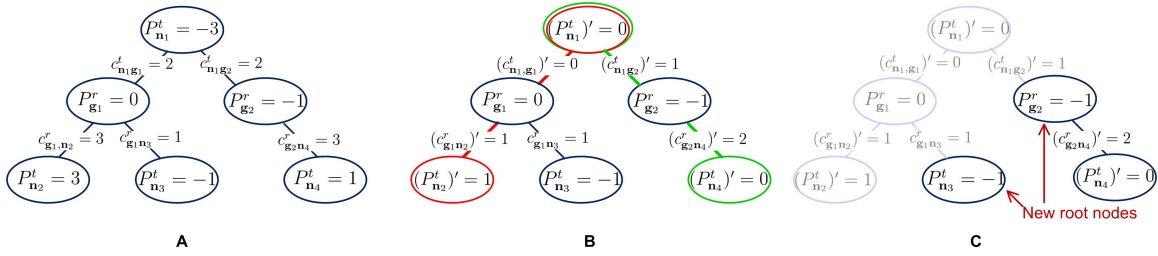}
\caption {Illustrative example of the ``pressure
transfer tree" and the corresponding operations. A) A tree generated in Step A and B;
B) Pressure transfer in Step C; C) Removal of depleted links and
neutralized roots in Step D.} \label{fig_tree}
\end{figure}

For notational convenience, denote $v^t_{nj}$, $v^r_{gk}$ and $c_{gk,nj}$ as  $v^t_{ {\mathbf{n}}}$, $v^r_{\mathbf{g}}$ and $c_{\mathbf{gn}}$, respectively, where $ {\mathbf{n}}=(n,j)$,
$\mathbf{g}=(g,k)$, $n,g\in\{1,...,G\}$, $j,k\in\{1,...,K\}$.
\begin{itemize}
\item{\bf Initialize the constraint assignment:} Randomly generalize
a \emph{constraint assignment policy}, i.e. $\{c^t_{\mathbf{ng}},c^r_{\mathbf{gn}}\}$
such that: $c^t_{\mathbf{ng}},c^r_{\mathbf{gn}}\in\mathbb{N}\cup\{0\}$, $
c^t_{\mathbf{ng}}+c^r_{\mathbf{gn}}=c_{\mathbf{gn}}$.
Calculate
$\{P^t_{\mathbf{n}},P^r_{\mathbf{g}}\}$:
$P^t_{\mathbf{n}}=v^t_{\mathbf{n}}-\sum_{\mathbf{g}\in\mathbb{S}}c^t_{\mathbf{ng}}$,
$P^r_{\mathbf{g}}=v^r_{\mathbf{g}}-\sum_{\mathbf{n}\in\mathbb{S}}c^r_{\mathbf{gn}}$.
\item{\bf Update the constraint assignment:} As illustrated in Fig.~\ref{fig_tree}, while there exist ``overloaded nodes", i.e. $P^t_{\mathbf{n}}<0$ or
$P^r_{\mathbf{g}}<0$, do the following to update $\{c^t_{\mathbf{gn}},c^r_{\mathbf{gn}}\}$:
\begin{itemize}\item{\bf A. Initialization:}
Select an ``overloaded node" with negative pressure. For instance, assume $P^t_{\mathbf{n}}<0$, we set $P^t_{\mathbf{n}}$
to be the root node of the ``pressure transfer tree", which is a
variation of the tree data structure, with its nodes storing the
pressures at the precoders and decorrelators, its link strengths storing the maximum
number of constraints that can be reallocated between the parent
nodes and the child nodes.
\item{\bf B. Add leaf nodes to the
pressure transfer tree:}

For every leaf node $P^x_{\mathbf{n}}$
($x\in\{t,r\}$):
\begin{itemize}
\item[] For every $\mathbf{g}$: If $c^{\overline{x}}_{\mathbf{ng}}>0$,
add $P^{\overline{x}}_{\mathbf{g}}$ as a child node of $P^x_{\mathbf{n}}$ with link
strength $c^{\overline{x}}_{\mathbf{ng}}$, where $\overline{x}$ is the
element in $\{t,r\}$ other than $x$.
\end{itemize}
\item{\bf C. Transfer pressure from root to leaf nodes:} For every leaf node with positive pressure, transfer
pressure from root to these leafs by updating the constraint
assignment policy $\{c^t_{\mathbf{gn}},c^r_{\mathbf{gn}}\}$. For instance, as
illustrated in Fig.~\ref{fig_tree}B,
$P^t_{\mathbf{n}_1}\xrightarrow{c^t_{\mathbf{n}_1\mathbf{g}_1}}P^r_{\mathbf{g}_1}\xrightarrow{c^r_{\mathbf{g}_1\mathbf{n}_2}}P^t_{\mathbf{n}_2}$
is a root-to-leaf branch of the tree (red lines). Update: $(c^t_{\mathbf{n}_1\mathbf{g}_1})'=
c^t_{\mathbf{n}_1\mathbf{g}_1}-\epsilon$, $(c^{r}_{\mathbf{g}_1\mathbf{n}_1})'=
c^{r}_{\mathbf{g}_1\mathbf{n}_1}+\epsilon$, $(c^r_{\mathbf{g}_1\mathbf{n}_2})'= c^r_{\mathbf{g}_1\mathbf{n}_2}-\epsilon$,
$(c^{t}_{\mathbf{n}_2\mathbf{g}_1})'= c^{t}_{\mathbf{n}_2\mathbf{g}_1}+\epsilon$. Hence we have
$(P^t_{\mathbf{n}_1})'=P^t_{\mathbf{n}_1}-\epsilon$ and
$(P^{t}_{\mathbf{n}_2})'=P^{t}_{\mathbf{n}_2}+\epsilon$, where $\epsilon =
\min\left(-P^t_{\mathbf{n}_1}, P^t_{\mathbf{n}_2}, c^t_{\mathbf{n}_1\mathbf{g}_1}, c^r_{\mathbf{g}_1\mathbf{n}_2}\right)$.
\item{\bf D. Remove the ``depleted" links and ``neutralized" roots:}
\begin{itemize}
\item If the strength of a link become 0 after Step C: Separate the
subtree rooted from the child node of this link from the original
pressure transfer tree.
\item If the root of a pressure transfer tree is nonnegative,
remove the root and hence the subtrees rooted from each child node
of the root become new trees. Repeat this process until all roots
are negative. For each newly generated pressure transfer tree,
repeat Steps B$\sim$D (Please refer to Fig.~\ref{fig_tree}C for an
example).
\end{itemize}
\item{\bf E. Exit Conditions:} Repeat Steps A$\sim$D until all
trees become empty (hence the network is IA feasible) or no new leaf
node can be added for any of the non-empty trees in Step B (hence the
network is IA infeasible). Exit the algorithm.~\hfill \IEEEQED
\end{itemize}
\end{itemize}

\subsection{Proof for Theorem~\ref{thm:inter_full}}
\label{pf_thm:inter_full} We shall first prove the optimality part
via the following two lemmas:
\begin{Lem2}The updated decorrelators $\{\mathbf{U}_{gk}\}$ in Step  2 of Algorithm~\ref{alg:inter_full} are the
optimal solution for problems \eqref{eqn:optuc2}.\label{lem:inter1}
\end{Lem2}
\proof Please refer to \cite{JafarD1} for the proof.\endproof

\begin{Lem2}The updated free elements in precoder $\{\mathbf{V}^F_{nj}\}$ in Step  3 of Algorithm~\ref{alg:inter_full} are the
optimal solution for problems \eqref{eqn:optvc2}.\label{lem:inter2}
\end{Lem2}
\proof Denote $\mathbf{Q}_{nj}=\sum_{g=1,\neq
n}^{G}\sum_{k=1}^KP_{nj}(\mathbf{U}^H_{gk}\mathbf{H}_{gk,n}
)^H$ $(\mathbf{U}^H_{gk}\mathbf{H}_{gk,n} )$. Note that
$\mathbf{Q}_{nj}$ is a positive semidefinite matrix, and we have in
\eqref{eqn:optvc2}:
\begin{eqnarray}
\nonumber &&\sum_{n=1\atop\neq
g}^{G}\sum_{j=1}^K\mbox{trace}\left((\mathbf{U}^H_{gk}\mathbf{H}_{gk,n}
\mathbf{V}^I_{nj})^H(\mathbf{U}^H_{gk}\mathbf{H}_{gk,n}
\mathbf{V}^I_{nj})\right)=\sum_{n=1\atop\neq
g}^{G}\sum_{j=1}^K\mbox{trace}\left((\mathbf{V}^I_{nj})^H\mathbf{Q}_{nj}
\mathbf{V}^I_{nj}\right)
\\\nonumber&=&\sum_{n=1\atop\neq
g}^{G}\sum_{j=1}^K
\mbox{trace}\left((\mathbf{Q}^{\frac{1}{2}}_{nj}\mathbf{V}^C_{nj}+
\mathbf{Q}^{\frac{1}{2}}_{nj}\mathbf{S}_{n}\mathbf{V}^F_{nj})^H(
\mathbf{Q}^{\frac{1}{2}}_{nj}\mathbf{V}^C_{nj}+
\mathbf{Q}^{\frac{1}{2}}_{nj}\mathbf{S}_{n}\mathbf{V}^F_{nj})\right)
\\&=&||\mathbf{Q}^{\frac{1}{2}}_{nj}\mathbf{V}^C_{nj}+
\mathbf{Q}^{\frac{1}{2}}_{nj}\mathbf{S}_{n}\mathbf{V}^F_{nj}||^2_F\label{eqn:fro}
\end{eqnarray}

By minimizing the Frobenius norm in \eqref{eqn:fro}, we have:
\begin{eqnarray}
\mathbf{V}^F_{nj}=-\left((\mathbf{Q}^{\frac{1}{2}}_{nj}\mathbf{S}_{n})^H(\mathbf{Q}^{\frac{1}{2}}_{nj}\mathbf{S}_{n})\right)^{-1}
(\mathbf{Q}^{\frac{1}{2}}_{nj}\mathbf{S}_{n})^H
\mathbf{Q}^{\frac{1}{2}}_{nj}\mathbf{V}^C_{nj}
=-(\mathbf{S}^H_{n}\mathbf{Q}_{nj}\mathbf{S}_{n})^{-1}\mathbf{S}^H_{n}
\mathbf{Q}_{nj}\mathbf{V}^C_{nj}\label{eqn:vf_sol}
\end{eqnarray}

\eqref{eqn:vf_sol} proofs the lemma.

Now we begin to prove the convergence part. Denote
\begin{eqnarray}I=\sum_{g=1}^{G}\sum_{n=1\atop\neq
g}^{G}\sum_{k=1}^K\sum_{j=1}^K\mbox{trace}\left((\mathbf{U}^H_{gk}\mathbf{H}_{gk,n}
\mathbf{V}^I_{nj})^H(\mathbf{U}^H_{gk}\mathbf{H}_{gk,n}
\mathbf{V}^I_{nj})\right)\end{eqnarray}

Then $I$ is non-negative. Moreover, from Lemma\ref{lem:inter1},
\ref{lem:inter2}, $I$ is non-increasing in each round of update.
Hence, following the analysis in \cite{JafarD1},
Algorithm~\ref{alg:inter_full} is surely to converge.

\subsection{Proof of Theorem~\ref{thm:drank}}\label{pf_thm:drank}
\begin{Lem2}\label{lem:vrank}$
\mbox{rank}\left(\left[\mathbf{V}^{I*}_{n1},\mathbf{V}^{I*}_{n2},...,
\mathbf{V}^{I*}_{nK}\right]\right)=\sum_{j=1}^Kd_{nj},
\;\forall n\in\{1,..,G\}$.
\end{Lem2}
\proof Denote the $q$-th column in
$\left[\mathbf{V}^{I*}_{n1},\mathbf{V}^{I*}_{n2},...,\mathbf{V}^{I*}_{nK}\right]$
and
$\left[\mathbf{V}^{C}_{n1},\mathbf{V}^{C}_{n2},...,\mathbf{V}^{C}_{nK}\right]$
as $\mathbf{v}^I_n(q)$ and $\mathbf{v}^C_n(q)$,
$q\in\mathbb{Q}=\{1,...,\sum_{j=1}^Kd_{nj}\}$,
respectively. From the intermediate precoder structure
\eqref{eqn:vform_1}, and the orthogonal constraint \eqref{eqn:orth}
(which means that all the $\mathbf{v}^C_n(q)$,
$q\in\{1,...,\sum_{j=1}^Kd_{nj}\}$ and the row
vectors in $\mathbf{S}_n$ are orthogonal to each other), we have
\begin{eqnarray}
\mathbf{v}^I_n(q) \in \mbox{span}(\mathbf{v}^C_n(q)) +
\mbox{span}(\mathbf{S}_n),\;\mathbf{v}^I_n(q)
\not\in\mbox{span}(\mathbf{S}_n)\label{eqn:vspan1}
\\\left(\mbox{span}(\mathbf{v}^C_n(q)) +
\mbox{span}(\mathbf{S}_n)\right) \cap \left(+_{p\neq
q,\in\mathbb{Q}}\left(\mbox{span}(\mathbf{v}^C_n(p)) +
\mbox{span}(\mathbf{S}_n)\right)\right)=\mbox{span}(\mathbf{S}_n)\label{eqn:vspan2}\end{eqnarray}
where $\mbox{span}(\mathbf{X})$ denotes the linear space spanned by
the column vectors of $\mathbf{X}$. From \eqref{eqn:vspan1},
\eqref{eqn:vspan2}, we have $\mathbf{v}^I_n(q)\not \in \mbox{span}(\{\mathbf{v}^I_n(p),p\neq
q,\in\mathbb{Q}\})$, $\forall q\in \mathbb{Q}$. This completes the proof.
\endproof
\begin{Lem2} Denote the $p$-th row of $\mathbf{U}^*_{nj}\mathbf{H}_{nj,n}$ as
$\mathbf{w}_{nj}(p)$, $p\in\{1,...,d_{nj}\}$. Then we have
$\mbox{span}(\mathbf{w}^H_{nj}(p))$ follows i.i.d. uniform
distribution in Grassmannian $\mathcal{G}(1,N^t_n)$.\label{lem:urank}
\end{Lem2}
\proof Denote $\mathbf{h}_{nj,n}(q)$ as the $q$-th row of
$\mathbf{H}_{nj,n}$, $q\in\{1,...,N^r_{nj}\}$. From
the i.i.d. fading assumption, $\mbox{span}(\mathbf{h}^H_{nj,n}(q))$
follows i.i.d. uniform distribution in $\mathcal{G}(1,N^t_n)$.
Since in \eqref{eqn:optvc2}, \eqref{eqn:optuc2}, the intra-cell
channel states $\{\mathbf{H}_{nj,n}\}$, $n\in\{1,...,G\}$,
$j\in\{1,...,K\}$ do not appear, $\{\mathbf{H}_{nj,n}\}$ are
independent of $\{\mathbf{V}^{I*}_{nj}\}$ and
$\{\mathbf{U}^*_{nj}\}$. Hence we have
$\mbox{span}(\mathbf{w}^H_{nj}(p))=\mbox{span}\left(\sum_{q=1}^{N^r_{nj}}u_{nj}(p,q)\mathbf{h}^H_{nj,n}(q)\right)
$ still uniformly distributed in $\mathcal{G}(1,N^t_n)$, where
$u_{nj}(p,q)$ is the element in the $p$-th row and $q$-th column of
$\mathbf{U}^*_{nj}$. Note that the rows in $\mathbf{U}^*_{nj}$ are
orthogonal to each other, $\mbox{span}(\mathbf{w}^H_{nj}(p))$ are
independent $p\in\{1,...,d_{nj}\}$. Moreover, since
$\mathbf{H}_{nj,n}$, $\mathbf{H}_{nj',n}$ are independent, $\mbox{span}(\mathbf{w}^H_{nj}(p))$ and
$\mbox{span}(\mathbf{w}^H_{nj'}(p'))$ are independent.
\endproof

From Lemma\ref{lem:urank}, we can easily deduce the following two
results
\begin{eqnarray}
\mbox{rank}\left(\left[\begin{array}{c}
\mathbf{H}^H_{n1,n}\mathbf{U}^*_{n1}
,\mathbf{H}^H_{n2,n}\mathbf{U}^*_{n2},...,
\mathbf{H}^H_{nK,n}\mathbf{U}^*_{nK}\end{array}
\right]
\right)=\sum_{j=1}^K d_{nj},\;\mbox{almost surely
(a.s.)}\label{eqn:urank}
\end{eqnarray}

and if $\sum_{j\in=1}^Kd_{nj}<N^t_g$,
\begin{eqnarray}\mathbf{v}\not\in\mbox{span}\left(\left[\begin{array}{c}
\mathbf{H}^H_{n1,n}\mathbf{U}^*_{n1}
,\mathbf{H}^H_{n2,n}\mathbf{U}^*_{n2},...,
\mathbf{H}^H_{nK,n}\mathbf{U}^*_{nK}\end{array}
\right]\right),\;\forall \mathbf{v}\in\mathbb{C}^{N^t_n\times
1}\mbox{ a.s. }\label{eqn:notv}
\end{eqnarray}

If $\sum_{j=1}^Kd_{nj}=N^t_n$, from
Lemma\ref{lem:vrank} and \eqref{eqn:urank},
\eqref{eqn:sufficient_rank} is proved. Otherwise, from
\eqref{eqn:notv}:
\begin{eqnarray}
&&\nonumber
\mbox{dim}\left(\mbox{span}\left(\left[\begin{array}{c}
\mathbf{H}^H_{n1,n}\mathbf{U}^*_{n1},...,
\mathbf{H}^H_{nK,n}\mathbf{U}^*_{nK}\end{array}
\right]\right) +
\left(\mbox{span}\left(\left[\mathbf{V}^{I*}_{n1},...,
\mathbf{V}^{I*}_{nK}\right] \right)\right)^\perp\right)
\\&&\nonumber
=\sum_{j=1}^Kd_{nj}+(N^t_n-\sum_{j=1}^Kd_{nj})=N^t_n,\;\mbox{a.s.}
\\&\Leftrightarrow&\nonumber
\mbox{dim}\left(\left(\mbox{span}\left(\left[\begin{array}{c}
\mathbf{H}^H_{n1,n}\mathbf{U}^*_{n1},...,
\mathbf{H}^H_{nK,n}\mathbf{U}^*_{nK}\end{array}
\right]\right)\right)^\perp \cap
\mbox{span}\left(\left[\mathbf{V}^{I*}_{n1},...,
\mathbf{V}^{I*}_{nK}\right] \right)\right)=0,\;\mbox{a.s.}
\\&\Leftrightarrow&\mathcal{N}\left(\left[\begin{array}{c}
\mathbf{H}^H_{n1,n}\mathbf{U}^*_{n1},...,
\mathbf{H}^H_{nK,n}\mathbf{U}^*_{nK}\end{array}
\right]
\right)\cap
\mathcal{R}\left(\left[\mathbf{V}^{I*}_{n1},\mathbf{V}^{I*}_{n2},...,
\mathbf{V}^{I*}_{nK}\right]\right)=\{0\},\;\mbox{a.s.}\label{eqn:RN}
\end{eqnarray}
where $\mathcal{N}(\mathbf{X})$ and $\mathcal{R}(\mathbf{X})$
denotes the null space and the range of matrix $\mathbf{X}$,
respectively. From \eqref{eqn:RN}, we have
\eqref{eqn:sufficient_rank}. This ends the proof.

\subsection{Proof of Theorem~\ref{thm:intra_full}}
\label{pf_thm:intra_full} We need to show:
A) Prove that the output of Algorithm~\ref{alg:intra_full}
$\{\mathbf{V}^*_{nj}\}$, $n\in\{1,...,G\}$, $j\in\{1,...,K\}$, is
indeed a solution for Problem~\ref{pro:intra}, i.e. it satisfies
constraints \eqref{eqn:viso}, \eqref{eqn:uvrank2} and
{\color{mblue}\eqref{eqn:spaceeq};} B) Show that the intra-cell interference power
\eqref{eqn:intra} under $\{\mathbf{V}^*_{nj}\}$ is 0.

We shall first prove the A) part. In Step  3, from the property of
SVD, we have:
\begin{eqnarray}
&&\mathbf{V}^{*H}_{nq}\mathbf{V}^*_{nq}=(\mathbf{A}'_{nq})^H\mathbf{A}'_{nq}=\mathbf{I}
\label{eqn:Aiso}
\\&&\mathbf{S}_{nq}=\left[\begin{array}{c}(\mathbf{S}'_{nq})^T,\;\mathbf{0}\end{array}\right]^T\label{eqn:Sshape}\end{eqnarray}
where $\mathbf{S}'_{nq}$ is a $d_{nq}\times d_{nq}$ diagonal
matrix.

From Theorem~\ref{thm:drank}, in Step  1, $\mathbf{L}_n(q)$ is full
rank almost surely. Note that $\mathbf{R}_n(q)$ is a lower
triangular matrix, in Step  2, we have:
\begin{eqnarray}
&&\mbox{rank}\left((\mathbf{U}^*_{nq})^H\mathbf{H}_{nq,n}\mathbf{V}'_{nq}\right)=
\mbox{rank}(\mathbf{L}'_{n}(q))=d_{nq},\;\forall
q\in\{1,...,N\}\mbox{ a.s.}\label{eqn:uvrank3}
\\&&(\mathbf{U}^*_{np})^H\mathbf{H}_{np,n}\mathbf{V}'_{nq}=\mathbf{0},\;\forall p\neq q\in\{1,...,N\}\label{eqn:intra_good}
\end{eqnarray}
where $\mathbf{L}'_{n}(q)$ is a $d_{nq}\times d_{nq}$ matrix
aggregated by the elements in the last $d_{nq}$ rows and columns
of $\mathbf{L}_{n}(q)$. From~\eqref{eqn:uvrank3}, we have
$\mbox{rank}(\mathbf{S}'_{nq})=\mbox{rank}(\mathbf{V}'_{nq})=d_{nq}$
almost surely. Hence,
$\mathbf{V}^*_{nq}=\mathbf{V}'_{nq}(\mathbf{S}'_{nq})^{-1}\mathbf{B}_{nq}$
almost surely. This result leads to the following equations:
\begin{eqnarray}
\nonumber
\mbox{rank}\left((\mathbf{U}^*_{nq})^H\mathbf{H}_{nq,n}\mathbf{V}^*_{nq}\right)&=&\mbox{rank}\left((\mathbf{U}^*_{nq})^H\mathbf{H}_{nq,n}\mathbf{V}'_{nq}(\mathbf{S}'_{nq})^{-1}\mathbf{B}_{nq}\right)
\\&=&\mbox{rank}\left((\mathbf{U}^*_{nq})^H\mathbf{H}_{nq,n}\mathbf{V}'_{nq}\right)
=d_{nq},\;\forall q\in\{1,...,N\}\mbox{ a.s.}\label{eqn:uvrank4}
\\\nonumber \left[\mathbf{V}^{*}_{n1},\mathbf{V}^{*}_{n2},...,
\mathbf{V}^{*}_{nK}\right]&=&\left[\mathbf{V}^{I*}_{n1},\mathbf{V}^{I*}_{n2},...,
\mathbf{V}^{I*}_{nK}\right]\cdot\left[\mathbf{Q}'_n(1)(\mathbf{S}'_{n1})^{-1}\mathbf{B}_{n1},\right.
\\&&\left.
\mathbf{Q}'_n(2)(\mathbf{S}'_{n2})^{-1}\mathbf{B}_{n2},...,\mathbf{Q}'_n(N)(\mathbf{S}'_{nK})^{-1}
\mathbf{B}_{nK}\right]\label{eqn:spaneeq}
\end{eqnarray}
\eqref{eqn:Aiso}, \eqref{eqn:uvrank4} and \eqref{eqn:spaneeq} proofs
the statement A).

Now we turn to the B) part. Statement B) is an immediate reference
of \eqref{eqn:intra_good}.

\subsection{Proof for Theorem~\ref{thm:equal2}}
\label{pf_thm:equal2}
{We need to prove statement  B) in Appendix\ref{pf_thm:equal}. Following the analysis in \cite[Appendix B]{J_Ruan}, after introducing the new transceiver structure in Definition~\ref{def:trans_partial}, the ``no more IA constraints than freedoms" constraint \eqref{eqn:feasible1} is extended to \eqref{eqn:feasible_p}. Moreover, constraints \eqref{eqn:orth_p}, \eqref{eqn:SR} and the fact that intra-cell channel states $\{\mathbf{H}_{gk,g}\}$ are independent of the inter-cell channel states $\{\mathbf{H}_{gk,n},n\neq g\}$, ensures that the statement in Theorem~\ref{thm:drank} still holds under partial connectivity. Hence, the solution set of Problem~\ref{pro:intra} is non-empty.} Substitute these solutions to Problem~\ref{pro:original}, then from \eqref{eqn:uvrank2}, \eqref{eqn:intra}, and \eqref{eqn:inter_1}, we have that $\{d^*_{nj}\}$, $\{\mathbf{V}^*_{nj}\}$,
$\{\mathbf{U}^*_{gk}\}$ must satisfy \eqref{eqn:rank_1} and \eqref{eqn:cs_cross_1}. This completes the  proof.

\subsection{Proof for Theorem~\ref{thm:DoF}}
\label{pf_thm:DoF} Due to the symmetry property of the system, in
Algorithm~\ref{alg:offline_p}, the stream assignment policy
$\{d^*_{gk}\}$, core space $\{\mathbf{V}^{C*}_{gk}\}$, free
space $\{\mathbf{S}^{t*}_{gk}\}$ and linear filter for decorrelator $\{\mathbf{U}^{r*}_{gk}\}$ hall be symmetrical $\forall
g\in\{1,...,G\}$. From $d_fK\le N^t$ and the first line of
\eqref{eqn:NT_a}, in Step 3 of Algorithm~\ref{alg:offline_p}, core
spaces assignment is feasible iff. $d_{gk}\le R_1$, $j\in\{1,...,K\}$.

From \eqref{eqn:feasible_p},
\eqref{eqn:d_t} and the second and third lines of \eqref{eqn:NT_a}, since the inter-cell partial connectivity state is the same for all users, in
Step 4 and 5 of Algorithm~\ref{alg:offline_p}, the number
of streams assigned to each MS $d_{gk}$, the dimension of the free spaces $S^t_{gk}$, and the dimension of the linear filter $S^r_{gk}$ shall be the same for all MSs. Moreover, as $\mathcal{N}^r_{gk,n}=\{0\}$, $\forall n,g,k$, we have that $S^r_{gk}$ shall be the maximum possible value, i.e. $N^r-R_1$. Denote $d=d_{gk}$, $S=S^t_{gk}$, then from the feasibility condition \eqref{eqn:feasible_p} and \eqref{eqn:NT_a}, stream assignment $\{d_{gk}=d\}$ is feasible if $d$ is achievable in the following problem:
\begin{eqnarray}
&&\max_{S}{d}\label{eqn:determineA}
\\&\mbox{S.t.}&\min(G-1,2J)K\min(d,\frac{R_2(d+S)}{N^t})\le S + N^r-d\label{eqn:feasible_sym}
\\\nonumber && d\in \{0,1,...,R_1\},\;\;S\in\{0,1,...,N^t-dK\}
\end{eqnarray}

By solving
\eqref{eqn:determineA}, we get \eqref{eqn:result1}.

\end{document}